\documentclass[conference]{IEEEtran}
\IEEEoverridecommandlockouts
\usepackage{cite}
\usepackage{amsmath,amssymb,amsfonts}
\usepackage{algorithmic}
\usepackage{graphicx}
\usepackage{textcomp}
\usepackage{xcolor}
\def\BibTeX{{\rm B\kern-.05em{\sc i\kern-.025em b}\kern-.08em
    T\kern-.1667em\lower.7ex\hbox{E}\kern-.125emX}}
    
\usepackage[braket]{qcircuit}
\usepackage{amsmath,amssymb,amsfonts}
\usepackage{graphicx}
\usepackage{xcolor}
\graphicspath{{svg-inkscape/}}
\usepackage{xparse}

\usepackage{tikz}
\usepackage{pgfplots}

\usetikzlibrary{patterns}
\usetikzlibrary{arrows}
\usetikzlibrary{external}

\pgfplotsset{{compat=1.14}}

\newcommand\plotscale{1}  
\newcommand\plottightscale{0.82}

\usepackage[final]{changes}  
\newcommand{\addcolor}{blue!70!green}
\setaddedmarkup{\textcolor{\addcolor}{\uwave{#1}}}
\setdeletedmarkup{\textcolor{red!70!black}{\sout{#1}}}

\makeatletter
\newcommand{\linebreakand}{%
  \end{@IEEEauthorhalign}
  \hfill\mbox{}\par
  \mbox{}\hfill\begin{@IEEEauthorhalign}
}
\makeatother
    
\begin{document}

\title{Exploiting Long-Distance Interactions \\ and Tolerating Atom Loss in \\ Neutral Atom Quantum Architectures\\
\thanks{This work is funded in part by EPiQC, an NSF Expedition in Computing, under grants CCF-1730449; in part by STAQ under grant NSF Phy-1818914; in part by DOE grants DE-SC0020289 and DE-SC0020331; and in part by NSF OMA-2016136 and the Q-NEXT DOE NQI Center. This research used resources of the Oak Ridge Leadership Computing Facility, which is a DOE Office of Science User Facility supported under Contract DE-AC05-00OR22725. Disclosure: F. Chong is also Chief Scientist at Super.tech and an advisor to Quantum Circuits, Inc.}
}

\author{\IEEEauthorblockN{Jonathan M. Baker}
\IEEEauthorblockA{\textit{Department of Computer Science} \\
\textit{University of Chicago}\\
Chicago, USA \\
jmbaker@uchicago.edu}
\and
\IEEEauthorblockN{Andrew Litteken}
\IEEEauthorblockA{\textit{Department of Computer Science} \\
\textit{University of Chicago}\\
Chicago, USA \\
litteken@uchicago.edu}
\and
\IEEEauthorblockN{Casey Duckering}
\IEEEauthorblockA{\textit{Department of Computer Science} \\
\textit{University of Chicago}\\
Chicago, USA \\
cduck@uchicago.edu}
\linebreakand
\IEEEauthorblockN{Henry Hoffmann}
\IEEEauthorblockA{\textit{Department of Computer Science} \\
\textit{University of Chicago}\\
Chicago, USA \\
hankhoffmann@cs.uchicago.edu}
\and
\IEEEauthorblockN{Hannes Bernien}
\IEEEauthorblockA{\textit{Pritzker School of Molecular Engineering} \\
\textit{University of Chicago}\\
Chicago, USA \\
bernien@uchicago.edu}
\and
\IEEEauthorblockN{Frederic T. Chong}
\IEEEauthorblockA{\textit{Department of Computer Science} \\
\textit{University of Chicago}\\
Chicago, USA \\
chong@cs.uchicago.edu}
}

\maketitle

\begin{abstract}
    Quantum technologies currently struggle to scale beyond moderate scale prototypes and are unable to execute even reasonably sized programs due to prohibitive gate error rates or coherence times. Many software approaches rely on heavy compiler optimization to squeeze extra value from noisy machines but are fundamentally limited by hardware. Alone, these software approaches help to maximize the use of available hardware but cannot overcome the inherent limitations posed by the underlying technology.
    
    An alternative approach is to explore the use of new, though potentially less developed, technology as a path towards scalability. In this work we evaluate the advantages and disadvantages of a Neutral Atom (NA) architecture. NA systems offer several promising advantages such as long range interactions and native multiqubit gates which reduce communication overhead, overall gate count, and depth for compiled programs. Long range interactions, however, impede parallelism with restriction zones surrounding interacting qubit pairs. We extend current compiler methods to maximize the benefit of these advantages and minimize the cost.

    Furthermore, atoms in an NA device have the possibility to randomly be lost over the course of program execution which is extremely detrimental to total program execution time as atom arrays are  slow to load. When the compiled program is no longer compatible with the underlying topology, we need a fast and efficient coping mechanism. We propose hardware and compiler methods to increase system resilience to atom loss dramatically reducing total computation time by circumventing complete reloads or full recompilation every cycle.
\end{abstract}


\section{Introduction}

\begin{figure}
    \centering
    \scalebox{0.8}{
        \def\svgwidth{1.25\columnwidth}
\begingroup%
  \makeatletter%
  \providecommand\color[2][]{%
    \errmessage{(Inkscape) Color is used for the text in Inkscape, but the package 'color.sty' is not loaded}%
    \renewcommand\color[2][]{}%
  }%
  \providecommand\transparent[1]{%
    \errmessage{(Inkscape) Transparency is used (non-zero) for the text in Inkscape, but the package 'transparent.sty' is not loaded}%
    \renewcommand\transparent[1]{}%
  }%
  \providecommand\rotatebox[2]{#2}%
  \newcommand*\fsize{\dimexpr\f@size pt\relax}%
  \newcommand*\lineheight[1]{\fontsize{\fsize}{#1\fsize}\selectfont}%
  \ifx\svgwidth\undefined%
    \setlength{\unitlength}{449.48490229bp}%
    \ifx\svgscale\undefined%
      \relax%
    \else%
      \setlength{\unitlength}{\unitlength * \real{\svgscale}}%
    \fi%
  \else%
    \setlength{\unitlength}{\svgwidth}%
  \fi%
  \global\let\svgwidth\undefined%
  \global\let\svgscale\undefined%
  \makeatother%
  \begin{picture}(1,0.31890019)%
    \lineheight{1}%
    \setlength\tabcolsep{0pt}%
    \put(0,0){\includegraphics[width=\unitlength,page=1]{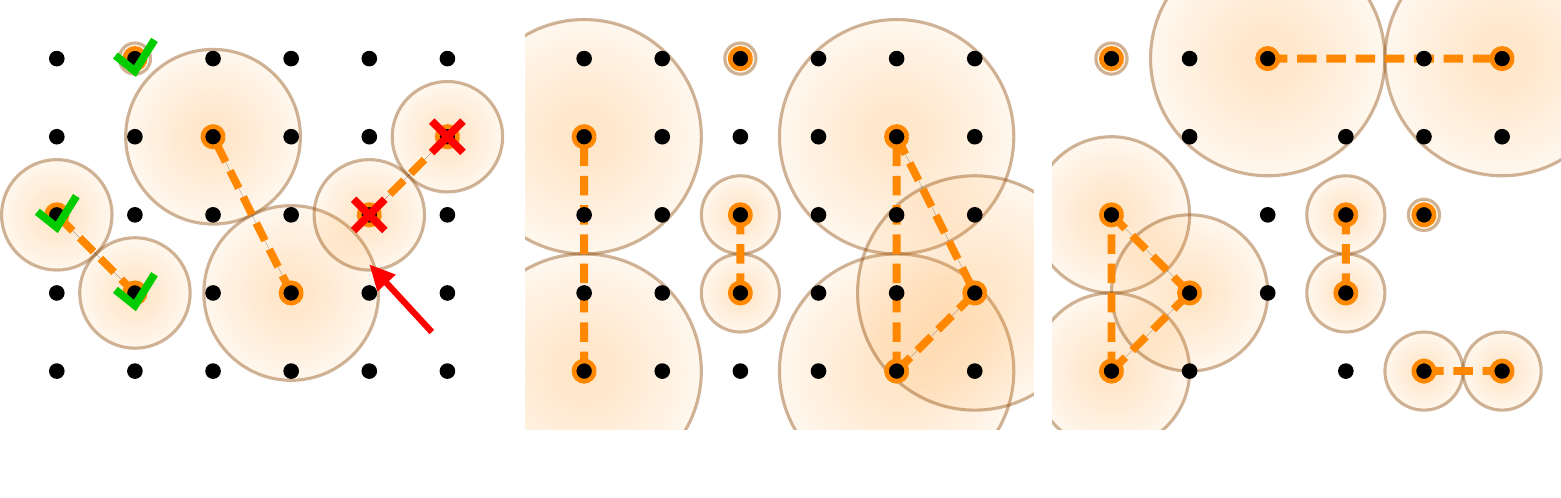}}%
    \put(0.16154025,0.00604207){\makebox(0,0)[t]{\lineheight{1.25}\smash{\begin{tabular}[t]{c}(a) Restriction Zone\end{tabular}}}}%
    \put(0.49942701,0.00604207){\makebox(0,0)[t]{\lineheight{1.25}\smash{\begin{tabular}[t]{c}(b) Max Distance 3\end{tabular}}}}%
    \put(0.83731378,0.00604207){\makebox(0,0)[t]{\lineheight{1.25}\smash{\begin{tabular}[t]{c}(c) Atom Loss\end{tabular}}}}%
  \end{picture}%
\endgroup%
    }
    \caption{Examples of interactions on a neutral atom device. (a) Interactions of various distances are permitted up to a maximum. Gates can occur in parallel if their zones do not intersect. The interaction marked with green checks can occur in parallel with the middle interaction. (b) The maximum interaction distance specifies which physical qubits can interact. Compiler strategies suited for this variable distance are needed for neutral atom architectures. (c) Neutral atom systems are prone to sporadic atom loss. Efficient adaptation to this loss reduces computation overhead.}
    \label{fig:intro-fig}
\end{figure}
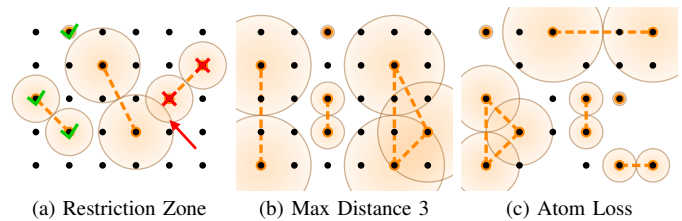

In the past several years, many leading gate-based quantum computing technologies such as trapped ions and superconducting qubits, have managed to build small-scale systems containing on the order of tens of qubits \cite{ionq, ibmq, bristlecone, rigetti}. However, each of these systems have unique scalability challenges \cite{oliver1, ion-scale}. For example, IBM devices have continued to grow in size while error rates have remained high, larger than what is needed for quantum error correction \cite{gottesman2010introduction}.
Trapped ion machines, despite many promising results, have fundamental challenges in controllability \cite{Murali-ISCA20}. It is unclear whether any of these platforms in present form will be capable of executing large-scale quantum computation needed for algorithms with quantum speedup like Grover's \cite{shor} or Shor's \cite{grover}. 

These challenges are fundamental to the underlying technology. Consequently, current approaches which aim to reduce error via software and push the limits of current devices are insufficient for long-term scalability. In recent years there has been a number of improvements to the compilation pipeline stretching across the hardware-software stack \cite{map1, map2, map3, routing1, routing2, scheduling1, cancel, xtalk}. Numerous studies have explored reductions in quantum circuit gate counts, depths, and communication costs via optimizations, but cannot overcome the fundamental scalability limitations of the technology.

An alternative approach is to consider emerging quantum technologies and new architectures. In this work, we explore hardware composed of arrays of individually-trapped, ultra-cold neutral atoms which have shown great promise and have unique properties which make them appealing from a software standpoint \cite{saffman}. These properties include: potential for high-fidelity quantum gates, indistinguishable qubits that enable scaling to many qubits, long-range qubit interactions that approximate global (or generally high) connectivity, and the potential to perform multiqubit ($\ge3$ operands) operations without expensive decompositions to native gates \cite{multiqubitgates}.

Neutral-atom (NA) architectures also face unique challenges. Long-range interactions induce zones of restriction around the operating qubits which prevent simultaneous operations on qubits in these zones. Most importantly, the atoms in a neutral atom device can be lost via random processes during and between computation. In the worst case, the compiled program no longer fits on the now sparser grid of qubits, requiring a reload of the entire array every cycle. This is a costly operation to repeat for thousands of trials. Coping with this loss in a time-efficient manner is important to minimizing the run time of input programs while not dramatically increasing either gate count or depth, both of which will reduce program success rate. In Figure \ref{fig:intro-fig} we show a small piece of a NA system with many gates of various sizes and distances being executed in parallel. Restriction zones are highlighted and importantly no pair of gates have intersecting restriction zones. When atoms are lost during computation, rather than a uniform grid we have a much sparser graph and qubits will be further apart on average. A key to the success of a NA system is resilience to loss of atoms, avoiding expensive reloads.

In this work, we explore the trade-offs in neutral atom architectures to assess both current viability and near future prospects. We extend current compilation methods to directly account for the unique NA constraints like long-range interactions, areas of restriction, and native implementation of multiqubit gates. We propose several coping strategies at the hardware and software level to adapt to loss of atoms during program execution; we evaluate the tradeoff of execution time and resilience to atom loss versus program success rate.

The field of quantum computation is still early in development and no clear winner for underlying hardware has emerged. It is vital to consider new technology and evaluate its potential early and often to determine viability as it scales. In this work, we do just that, considering a neutral atom architecture of size comparable to target hardware sizes for competing technologies. Despite comparably higher gate errors and lack of large scale demonstration, we determine if the unique properties offered by this new technology enable more efficient computation at scale. Perhaps more importantly, this work can serve as a guide for hardware developers. We demonstrate that the fundamental problem of atom loss can be mitigated via software solutions and doesn't need to be highly optimized at the hardware level. This allows hardware engineers to focus on other fundamental problems which cannot easily be mitigated by software, such as gate error rate.

In this work we introduce a scalable neutral atom architecture based on demonstrated physical implementations which permit long range interactions and native multiqubit gates. The specific major contributions of our work are the following:
\begin{itemize}
    \item Adapt current quantum compiler technology by extending prior work to explicitly account for interaction distance, induced restriction zones, and multiqubit gates.
    \item Evaluate system-wide implications of these properties, specifically reduced gate counts and depth at the cost of increased serialization. Our compiler exploits the gain while mitigating this cost.
    \item Demonstrate, via simulation based on experimental results and through program error analysis, the ability of NA systems to quickly surpass competitors in the intermediate-term despite currently worse gate errors.
    \item Model sporadic atom loss in NA systems and propose hardware and compiler solutions to mitigate run time and program error rate overheads. We explore each strategy's resilience to atom loss, the effect on expected program success rate, and overall run time.
\end{itemize}

\section{Background}

\begin{figure}
    \centering
    \def\svgwidth{\columnwidth}
    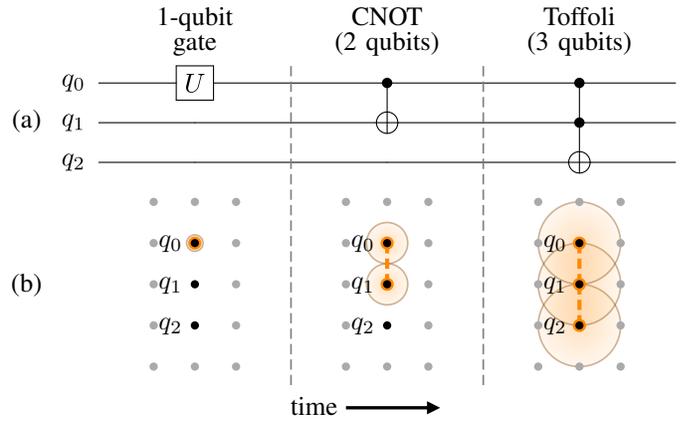
    \caption{A quantum circuit with a 1, 2, and 3 qubit gate translated to interactions on a NA device. These systems allow the execution of multiqubit gates. For 2 and 3 qubit gates the interacting qubits are excited to Rydberg states. Interactions are possible if all interacting qubits are closer than the maximum interaction distance.}
    \label{fig:quantum-circuit-gates}
\end{figure}

\subsection{Quantum Computation and the Gate Model}
The fundamental unit of quantum computing is the quantum bit, or qubit, which exists as a linear superposition between the $\ket{0}$ and the $\ket{1}$ states. Most quantum programs manipulate a register of $N$ quantum bits where the state of the qubits is given as a linear superposition of the $2^N$ basis vectors. This state evolves by the application of operations or gates. For example, the single qubit $X$ gate transforms $\ket{\psi} = \alpha\ket{0} + \beta\ket{1}$ into $X\ket{\psi} = \beta\ket{0} + \alpha\ket{1}$. Gates can also operate on multiple qubits at a time. For example, the CNOT gate applies an $X$ gate to the target qubit if and only if the control qubit is in the $\ket{1}$ state. These multiqubit operators are able to produce entanglement between the qubit states. Together, superposition and entanglement are central to the expected power of quantum computation. For a complete introduction see \cite{mikeike}.

Most quantum programs for gate-based quantum computation are expressed in the quantum circuit model. On most hardware platforms, only single and two qubit gates are supported, requiring complex operations like the generalized Toffoli, a common subcircuit in many quantum algorithms, to be decomposed into smaller pieces. Furthermore, most hardware only supports a small highly calibrated universal set of gates, requiring input programs be rewritten in terms of these gates. A small piece of a circuit is found in Figure \ref{fig:quantum-circuit-gates}a.

Two important metrics for quantum programs are the depth of the quantum program, given as the length of the longest critical path from inputs to outputs, and the gate count, that is how many operations it takes to perform the desired algorithm. Both are important for near-term quantum computation which is noise prone. Qubits have limited coherence time, likely erasing a qubit's information by a time limit. Gate error rates are fairly high so computations with many multiqubit gates are less likely to succeed.

\subsection{The Quantum Compilation Problem}
In order to maximize the probability of program success, quantum circuits often undergo extensive compilation and optimization. Compilation generally falls into two main categories: circuit optimization to minimize total number of gates, and translation of input programs to fit the constraints of the target hardware. The latter is often the focus of near-term compilation strategies which break the problem into three main steps: mapping, routing, and scheduling. 

In mapping, the program qubits must be assigned to hardware qubits with the goal of minimizing the distance between interacting qubits over the course of the program. As noted, most hardware only supports interactions between a limited set of qubits. Qubits mapped too far from each other must be moved nearby by inserting SWAPs or other communication operations before interacting. This communication is often very expensive and every extra gate needed for communication contributes to the overall error rate of the final program. It is common for the mapping and routing steps to occur in tandem as routing changes the mapping of the qubits over the course of the program. Finally, scheduling consists of deciding when to execute which gates and is usually dictated by factors such as run time or crosstalk where we may delay gates to avoid crosstalk effects but possibly increase runtime.

\subsection{Neutral Atoms}
We want to briefly introduce some background on the underlying neutral atom technology.  A nice introduction can be found in \cite{Henriet_2020}. Atoms in a NA system are trapped via reconfigurable, optical tweezer arrays. These atoms can be arranged in one, two, or even three dimensions \cite{na2, na3, na4, na5}. We consider regular square 2D geometries in this work, but arbitrary arrangements of atoms are possible. Historically, one of the major difficulties with scalable neutral atom systems was the probabilistic nature of atom trapping, but this challenge has since been overcome and defect-free arrays of more than 100 atoms \cite{Ohl_de_Mello_2019} have been demonstrated. The loading of qubits into the array is relatively slow, on the order of one second, compared to program execution which usually takes milliseconds. 

The single atom qubit states can be manipulated using Raman transitions which implement single qubit gates. In order to execute gates between qubits, atoms are optically coupled to highly excited Rydberg states leading to a strong dipole-dipole interaction between the atoms \cite{na-gates}. These Rydberg interactions enable multiple atoms at once to interact strongly and are used to realize multiqubit gates \cite{multiqubitgates}. Furthermore, due to the long range of these interactions, gates between qubits which are not directly adjacent in the atom array are feasible. Importantly, these interactions induce a zone of restriction as a function of the distance. Two gates can only occur in parallel if their restriction zones do not overlap.

\section{Neutral Atom Compiler and Methodology}

\subsection{Mapping, Routing, and Scheduling}
In this work we focus on adapting currently available and effective compilation methods \cite{look1, look2, look4, qiskit} to directly account for the unique properties of neutral atom architectures. We focus on mapping, routing and scheduling of the quantum compilation problem. Other optimizations, such as circuit synthesis, gate optimization, or even pulse optimization, can be performed as well, but are not the focus of this work. Many of the primary advantages and disadvantages of the neutral atom hardware can be reduced to modifications of the hardware topology or interaction model given to the compiler. 

We represent the underlying topology as a graph, where nodes are hardware qubits and edges are between nodes which can interact. We model the underlying hardware as a 2D grid of qubits and, for a given instance, we fix the maximum interaction distance $d_{max}$. Therefore, there is an edge between nodes $u, v$ if $d(u, v) \le d_{max}$. We model the restriction zones as circles of radius $r$ centered at each of the interacting qubits. In this work we model this radius as $f(d) = d / 2$ where $d$ is the maximum distance between interacting qubits, pairwise. We have ensured this generalizes to any number of qubits in order to support multiqubit gates. In practice, devices may require a different function of $d$. The larger this radius the fewer possible parallel interactions can occur.

For most quantum programs, the entire control flow is known at compile time making optimal solutions for mapping and routing possible but exponentially difficult to find.  Therefore the dominant solutions are heuristics. We have extended prior work on lookahead heuristics.  Lookahead bases mapping and routing on the sum of weighted future interactions with operations further into the future weighted less. The entire circuit is mapped and routed in steps. At each step, we consider the \textit{weighted interaction graph} where nodes are \textit{program qubits} and edges between nodes are weighted by the lookahead function:
$$w(u, v) = \sum_{\ell \ge \ell_{c}} e^{-|\ell_c - \ell|}$$
where $w(u, v)$ is the weight between program qubits $u$ and $v$, $\ell$ is a layer of the program ,and $\ell_{c}$ is the current layer, i.e. the frontier of the program DAG. When considering a multiqubit gate we add this weighting function between all pairs of qubits in the gate.

For the initial mapping, we begin by placing the qubits with the greatest interaction weight in the weighted interaction graph. We place these qubits adjacent in the center of the device. For every subsequent qubit in this graph we consider all possible assignments to hardware qubits and choose the best based on a score:
$$s(u, h) = \sum_{\text{mapped}\ v} d(h, \varphi(v)) \times w(u, v)$$
where $h$ is the potential hardware location, and $\varphi$ is the mapping from program qubits to hardware qubits. The goal is to place qubits which interact frequently close to each other in order to avoid extra SWAPs during routing. We choose the hardware location $h$ which minimizes this score. We place qubits ordered by their weight to those previously mapped, greatest first.

For routing and scheduling, we proceed layer by layer, considering operations in the frontier as potential gates to execute. Ideally, we would execute all operations in the frontier in parallel, however if interacting qubits are not close enough or the zones of restriction intersect, this isn't possible. Instead, we first select to execute operations in the frontier which do not have intersecting zones. For any remaining long distance operations in the frontier, we compute the best set of SWAPs to move qubits within the interaction distance. We want to select a path of SWAPs with two goals in mind: the shortest path and the least disruptive to future interactions. This leads to the following scoring function:
\begin{align*}
    s(u, h) = \sum_{v} &[d(\varphi(u), \varphi(v)) - d(h, \varphi(v))] \times w(u, v) + \\ &[d(h, \varphi(v)) - d(\varphi(u), \varphi(v))] \times w(\varphi^{-1}(h), v)
\end{align*}
where $h$ is the new location for $u$ after the SWAP. We choose the $h$ which maximizes this function but is also strictly closer to the most immediate interaction for $u$ and $v$. In this function, moving further away from future interactions or displacing the qubit in position $h$ by moving it far from its future interactions is penalized. This guarantees the qubit always moves closer to its target. The SWAP is executed if it can run parallel with the other executable operations, otherwise we must wait. We proceed until all operations have been executed. Our compiler and evaluation source code is available at \cite{github-link}.

To scale target programs up to hundreds to thousands of qubits, the heuristics used are fairly simple and fast.
One clear advantage of NA is that simpler and faster heuristics will suffice in practice because large interaction distances make the topology densely connected thus saving communication cost. 

We validated our compiler by compiling small programs via IBM's Qiskit compiler \cite{qiskit} with lookahead enabled against our compiler with maximum interaction distance (MID) set to 1 and no restriction zones for two benchmarks, one parallel and one not. In both cases, our compiler closely matched Qiskit in both gate count and depth.

\subsection{Benchmarks}
For this work, we have chosen a set of quantum programs which are parametrized, the input size can be specified, to allow us to study how the advantages and disadvantages of a NA system change as the program size increases. Specifically, we study Bernstein-Vazirani \cite{bv}, a common quantum benchmark, with the all 1s oracle to maximize gates, Cuccaro Adder \cite{cuccaro}, a ripple carry adder with no parallelism, the CNU gate \cite{barenco}, a logarithmic depth and highly parallel decomposition of a very common subcircuit, QFT Adder \cite{qft-adder}, a circuit with two QFT components and a highly parallel addition component, and QAOA for MAX-CUT \cite{qaoa}, a promising near-term algorithm, on random graphs with a fixed edge density of 0.1.

\subsection{Experimental Setup}
For most experiments, we compile our benchmarks, with sizes up to 100, on a $10 \times 10$ NA device. We have a fixed radius of restriction but vary max interaction distance from 1 (emulating superconducting systems) up to the maximum needed for global connectivity (here $hypot(9, 9)\approx13$). In relevant benchmarks we compile with decomposed multiqubit gates and without. All experiments were run using on a machine using Python 3.7 \cite{python}, Intel(R) Xeon(R) Silver 4110 2.10GHz, 132 GB of RAM, on Ubuntu 16.04 LTS.  All plot error bars show $\pm1$ standard deviation.
\section{Unique Advantages of Neutral Atom Architectures}
In this section, we explore promising architectural advantages provided by the neutral atom technology. We examine long range interactions where atoms distant on the device can interact similar to a device with high connectivity. However, the cost of this longer range interaction is higher serialization due to the proportional increase in restricted area. Second, we explore the native execution of multiqubit gates on the NA platform. Since NA technology is still in its early stages, it can be unfair to compare expected program success rates from current gate error rates and coherence times. We analyze common metrics, gate count and depth, which are good predictors of a program's success rate if executed.

\subsection{Long Range Interactions}
\begin{figure*}[h]
    \centering
    \scalebox{\plotscale}{%
    \makebox[1\textwidth][c]{%
        \quad\quad\quad\quad%
%
%
%
%
\begin{tikzpicture}[baseline,scale=1,trim axis left,trim axis right]
\pgfplotsset{every tick label/.append style={font=\small}}
\pgfplotsset{every axis label/.append style={font=\small}}

    \begin{axis}[
        name=plot0,
        title={Gate Count Savings from Interaction Distance},
        xlabel={},
        ylabel={reduction in gate count},
        symbolic x coords={BV,CNU,Cuccaro,QFT-Adder,QAOA},
        width={\columnwidth},
        height={0.5*\columnwidth},
        ybar={2pt},
        bar width={4pt},
        enlargelimits=0.125,
        ymin=0, ymax=109,
        xtick=data,
        ,
        legend style={draw=none, fill=none, at={(0.5,1.03)},anchor=north,font=\small},
        legend columns=-1,
        legend image code/.code={\draw[#1, draw=none] (0em,-0.2em) rectangle (0.6em,0.4em);},
        axis line style={draw=black!20!white},
        axis on top,
        y axis line style={draw=none},
        axis x line*=bottom,
        tick style={draw=none},
        yticklabel={\pgfmathparse{\tick*1}\pgfmathprintnumber{\pgfmathresult}\%},
        clip=false,
        enlarge y limits=0,
        ,
        x tick label style={},
        grid=none,
        ymajorgrids=false,
        ,
        ,
        nodes near coords always on top/.style={
            every node near coord/.append style={
                anchor=south,
                rotate=0,
                font=\small,
                inner sep=0.2em,
            },
        },
        nodes near coords always on top,
    ]
        \addplot[
            style={
                color=transparent,
                draw=none,
                fill={rgb,555:red,255;green,113;blue,0},
                ,
                mark=none,
                ,
                ,
            },
            error bars/.cd,y dir=both,y explicit]
        coordinates {
            (BV, 47.92381413833125) +- (0, 11.677574155643988)
            (CNU, 21.041876739755455) +- (0, 10.249325344081422)
            (Cuccaro, 39.6860304246132) +- (0, 2.9439452870117964)
            (QFT-Adder, 19.791352039712592) +- (0, 5.882759334171374)
            (QAOA, 25.44873191377327) +- (0, 13.522706614837526)
        };
        \addlegendentry{2~~~~};

        \addplot[
            style={
                color=transparent,
                draw=none,
                fill={rgb,255:red,209;green,216;blue,0},
                ,
                mark=none,
                ,
                ,
            },
            error bars/.cd,y dir=both,y explicit]
        coordinates {
            (BV, 58.81005636542877) +- (0, 15.470094974543148)
            (CNU, 41.08646625363292) +- (0, 4.2957804752331095)
            (Cuccaro, 42.16529641851084) +- (0, 4.1643639360828555)
            (QFT-Adder, 31.45544857030979) +- (0, 4.912528782856494)
            (QAOA, 36.7460419515834) +- (0, 21.921610525965345)
        };
        \addlegendentry{3~~~~};

        \addplot[
            style={
                color=transparent,
                draw=none,
                fill={rgb,555:red,209;green,216;blue,0},
                ,
                mark=none,
                ,
                ,
            },
            error bars/.cd,y dir=both,y explicit]
        coordinates {
            (BV, 62.92173235327077) +- (0, 17.201251174530647)
            (CNU, 46.64451595900118) +- (0, 5.464658411784565)
            (Cuccaro, 42.596787119842176) +- (0, 4.087264490828495)
            (QFT-Adder, 38.63594952104611) +- (0, 3.8584949707434166)
            (QAOA, 42.00972636501451) +- (0, 25.98588232211303)
        };
        \addlegendentry{4~~~~};

        \addplot[
            style={
                color=transparent,
                draw=none,
                fill={rgb,255:red,0;green,205;blue,110},
                ,
                mark=none,
                ,
                ,
            },
            error bars/.cd,y dir=both,y explicit]
        coordinates {
            (BV, 63.43279929745908) +- (0, 17.723724437979904)
            (CNU, 48.22394256048574) +- (0, 6.8193125838287685)
            (Cuccaro, 42.66604957940502) +- (0, 4.151440964623516)
            (QFT-Adder, 40.77827252376687) +- (0, 4.558183347141504)
            (QAOA, 43.78436213303857) +- (0, 27.483347989351984)
        };
        \addlegendentry{5~~~~};

        \addplot[
            style={
                color=transparent,
                draw=none,
                fill={rgb,555:red,0;green,205;blue,110},
                ,
                mark=none,
                ,
                ,
            },
            error bars/.cd,y dir=both,y explicit]
        coordinates {
            (BV, 64.83616878700565) +- (0, 18.13303644670697)
            (CNU, 48.689178570357655) +- (0, 7.076828429161466)
            (Cuccaro, 42.732293510466505) +- (0, 4.228534223300321)
            (QFT-Adder, 41.46360602337835) +- (0, 4.884318707632983)
            (QAOA, 44.742599652510414) +- (0, 28.350934791514337)
        };
        \addlegendentry{8~~~~};

        \addplot[
            style={
                color=transparent,
                draw=none,
                fill={rgb,255:red,125;green,0;blue,255},
                ,
                mark=none,
                ,
                ,
            },
            error bars/.cd,y dir=both,y explicit]
        coordinates {
            (BV, 65.64240941083486) +- (0, 18.528102023893993)
            (CNU, 48.689178570357655) +- (0, 7.076828429161466)
            (Cuccaro, 42.74106851924151) +- (0, 4.239686276022213)
            (QFT-Adder, 41.46360602337835) +- (0, 4.884318707632983)
            (QAOA, 44.742599652510414) +- (0, 28.350934791514337)
        };
        \addlegendentry{13};
    \end{axis}

\end{tikzpicture}%
        \hfill\quad\quad\quad\quad%
%
%
%
%
\begin{tikzpicture}[baseline,scale=1,trim axis left,trim axis right]
\pgfplotsset{every tick label/.append style={font=\small}}
\pgfplotsset{every axis label/.append style={font=\small}}

    \begin{axis}[
        name=plot8,
        title={BV Gate Count},
        xlabel={maximum interaction distance},
        ylabel={post-compilation gate count},
        width={0.8*\columnwidth},
        height={0.5*\columnwidth},
        xmin=1, xmax=13, ymin=-1, ymax=1308,
        ,
        legend style={
            draw=none,
            at={(1,0.5)},
            anchor=east,
            font=\small},
        ,
        clip=false,
        axis line style={draw=none},
        tick style={draw=none},,
        clip, extra x ticks={1, 13}, legend style={at={(1.1,0.5)},anchor=west},
    ]
        \addplot[color={rgb,255:red,255;green,113;blue,0}, ]
            table[x=maximum-interaction-distance, y=99 post-compilation-gate-count 0, col sep=comma]
            {data/mid-vs-gc-bv-gate-count.csv}
        ;
        \addlegendentry{99};

        \addplot[color={rgb,555:red,255;green,113;blue,0}, ]
            table[x=maximum-interaction-distance, y=87 post-compilation-gate-count 1, col sep=comma]
            {data/mid-vs-gc-bv-gate-count.csv}
        ;
        \addlegendentry{87};

        \addplot[color={rgb,255:red,209;green,216;blue,0}, ]
            table[x=maximum-interaction-distance, y=75 post-compilation-gate-count 2, col sep=comma]
            {data/mid-vs-gc-bv-gate-count.csv}
        ;
        \addlegendentry{75};

        \addplot[color={rgb,555:red,209;green,216;blue,0}, ]
            table[x=maximum-interaction-distance, y=63 post-compilation-gate-count 3, col sep=comma]
            {data/mid-vs-gc-bv-gate-count.csv}
        ;
        \addlegendentry{63};

        \addplot[color={rgb,255:red,0;green,205;blue,110}, ]
            table[x=maximum-interaction-distance, y=51 post-compilation-gate-count 4, col sep=comma]
            {data/mid-vs-gc-bv-gate-count.csv}
        ;
        \addlegendentry{51};

        \addplot[color={rgb,555:red,0;green,205;blue,110}, ]
            table[x=maximum-interaction-distance, y=39 post-compilation-gate-count 5, col sep=comma]
            {data/mid-vs-gc-bv-gate-count.csv}
        ;
        \addlegendentry{39};

        \addplot[color={rgb,255:red,125;green,0;blue,255}, ]
            table[x=maximum-interaction-distance, y=27 post-compilation-gate-count 6, col sep=comma]
            {data/mid-vs-gc-bv-gate-count.csv}
        ;
        \addlegendentry{27};

        \addplot[color={rgb,555:red,125;green,0;blue,255}, ]
            table[x=maximum-interaction-distance, y=15 post-compilation-gate-count 7, col sep=comma]
            {data/mid-vs-gc-bv-gate-count.csv}
        ;
        \addlegendentry{15};

        \addplot[color={rgb,255:red,255;green,113;blue,0}, ]
            table[x=maximum-interaction-distance, y=3 post-compilation-gate-count 8, col sep=comma]
            {data/mid-vs-gc-bv-gate-count.csv}
        ;
        \addlegendentry{3};

    \end{axis}

\end{tikzpicture}
        \quad\quad\quad\quad\hfill%
    }}
    \caption{Post-compilation gate count across benchmarks. On the left are percent savings over the distance 1 baseline averaged over program sizes up to 100 qubits.  Each color is a max interaction distance.  Noticeably, there is less additional improvement as the MID increases, indicating most benefit is gained for smaller distances. On the right is a sample benchmark (holds in general) with many program sizes compiled for the whole range of MIDs. As the program size increases, larger MID show benefit before flattening off.}
    \label{fig:mid-v-gc}
\end{figure*}
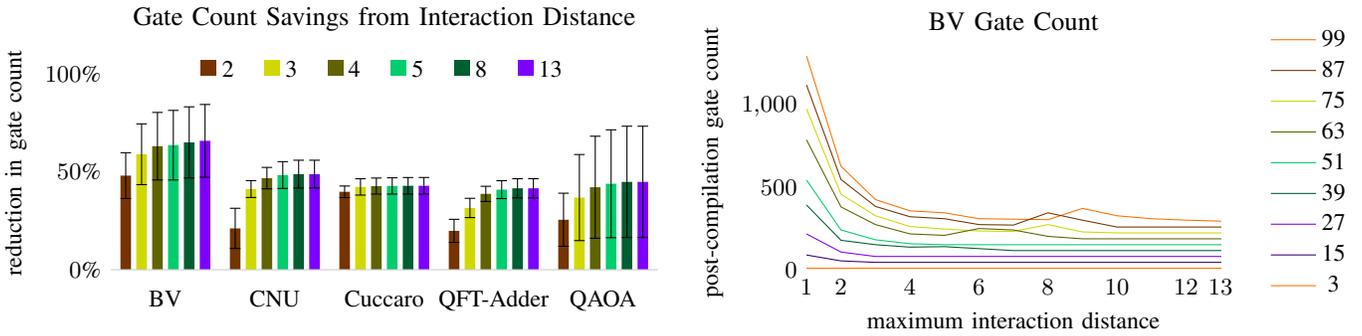

Trapped ion and superconducting architectures currently support qubit interaction only between \textit{adjacent} qubits. In SC systems this usually corresponds to a 2D grid or some other sparse connectivity, where each qubit is able to interact with a small number of qubits. One of the important promises of trapped ions is all-to-all connectivity where each qubit can interact freely with any other qubit in the same trap. Each trap however, is currently limited by the number of ions it can support and expensive interactions across different traps.

In NA architectures, the connectivity lies somewhere between these two extremes. The atoms, while often arranged in a 2D grid, have all-to-all connectivity beyond immediate neighbors, i.e. within a fixed radius. This radius is dictated by the capabilities of the hardware and can theoretically reach as large as the device.  However, current demonstrations have been more limited, for example up to distance 4. In this work, our experiments analyze the full sweep of interaction distances to understand the importantance of long range interactions to optimizing program success rate predictors.

Long range interactions in NA are not free, we define an area of restriction imposed by interacting qubits at a distance $d$ from each other, $f(d)$. Specifically, given this interaction distance between qubits of the set $Q$ all other qubits $q \not\in Q$ with distance less than $f(d)$ to \textit{any} of the interacting qubits cannot be operated on in parallel. Furthermore, suppose we have two operations to be performed in parallel. These two operations can only execute in parallel if their areas of restriction do not overlap. For experiments in this work we explore the function $f(d) = d / 2$. Intuitively, as this function becomes more restrictive, i.e. the areas surrounding the interacting qubits get larger, fewer total operations will be executable in parallel, affecting the total execution time of the program.

Long range interactions are important for reducing the total number of gates required for execution on devices with relatively limited connectivity. Limited connectivity requires compilers to add in many extra SWAP operations. The lower the connectivity, the greater the average distance between qubits on the device therefore more SWAPs are required to execute multiqubit gates between arbitrary sets of qubits. In Figure \ref{fig:mid-v-gc}, we explore the gate counts of compiled programs for various sizes over a range of maximum interaction distances up to the largest possible distance supported on the device. In each of these experiments, all programs are compiled to 1 and 2 qubit gates only.  Intuitively, we might assume having a larger maximum interaction distance will necessarily be better than a smaller one since it emulates global connectivity therefore not requiring any additional SWAP operations. In general, we find the most benefit in the first few improvements in max interaction distance with more relative gain for larger programs. The reduction in gate count is due solely to a reduction in total SWAPs.

Importantly, the benefit obtained from increasing max interaction distance tapers off with vanishing benefit. The rightmost points in these figures correspond to an interaction distance the full width of the device, providing all-to-all connectivity. At this distance no additional SWAP gates are required, so this is the minimum possible number of gates to execute the input program.  This distance is not required to obtain the minimum (or near the minimum). In fact, a smaller interaction distance is sufficient. This is promising in cases where large interaction distances cannot be obtained and hardware engineers can focus on building higher fidelity short to mid range interactions. For larger devices, the curves will be similar, however, requiring increasingly larger interaction distances to obtain the minimum. The shape of the curve will be more elongated, related directly to the average distance between qubits.

\begin{figure*}[h]
    \centering
    \scalebox{\plotscale}{%
    \makebox[1\textwidth][c]{%
        \quad\quad\quad\quad%
%
%
%
%
\begin{tikzpicture}[baseline,scale=1,trim axis left,trim axis right]
\pgfplotsset{every tick label/.append style={font=\small}}
\pgfplotsset{every axis label/.append style={font=\small}}

    \begin{axis}[
        name=plot0,
        title={Depth Savings from Interaction Distance},
        xlabel={},
        ylabel={reduction in depth},
        symbolic x coords={BV,CNU,Cuccaro,QFT-Adder,QAOA},
        width={\columnwidth},
        height={0.5*\columnwidth},
        ybar={2pt},
        bar width={4pt},
        enlargelimits=0.125,
        ymin=0, ymax=109,
        xtick=data,
        ,
        legend style={draw=none, fill=none, at={(0.5,1.03)},anchor=north,font=\small},
        legend columns=-1,
        legend image code/.code={\draw[#1, draw=none] (0em,-0.2em) rectangle (0.6em,0.4em);},
        axis line style={draw=black!20!white},
        axis on top,
        y axis line style={draw=none},
        axis x line*=bottom,
        tick style={draw=none},
        yticklabel={\pgfmathparse{\tick*1}\pgfmathprintnumber{\pgfmathresult}\%},
        clip=false,
        enlarge y limits=0,
        ,
        x tick label style={},
        grid=none,
        ymajorgrids=false,
        ,
        ,
        nodes near coords always on top/.style={
            every node near coord/.append style={
                anchor=south,
                rotate=0,
                font=\small,
                inner sep=0.2em,
            },
        },
        nodes near coords always on top,
    ]
        \addplot[
            style={
                color=transparent,
                draw=none,
                fill={rgb,555:red,255;green,113;blue,0},
                ,
                mark=none,
                ,
                ,
            },
            error bars/.cd,y dir=both,y explicit]
        coordinates {
            (BV, 57.08582816140426) +- (0, 14.043914658978116)
            (CNU, 7.926223278874841) +- (0, 23.81790188240674)
            (Cuccaro, 32.158712828932714) +- (0, 3.2402793223108652)
            (QFT-Adder, 21.811495218863627) +- (0, 9.142317781751204)
            (QAOA, 24.529471577233345) +- (0, 12.728842171547866)
        };
        \addlegendentry{2~~~~};

        \addplot[
            style={
                color=transparent,
                draw=none,
                fill={rgb,255:red,209;green,216;blue,0},
                ,
                mark=none,
                ,
                ,
            },
            error bars/.cd,y dir=both,y explicit]
        coordinates {
            (BV, 67.10060093961656) +- (0, 17.361543628197627)
            (CNU, 35.64610608462867) +- (0, 11.656780324484814)
            (Cuccaro, 30.467031409532684) +- (0, 4.632027259154623)
            (QFT-Adder, 21.34113060911478) +- (0, 6.500581567170287)
            (QAOA, 29.946326378000812) +- (0, 26.07029596831074)
        };
        \addlegendentry{3~~~~};

        \addplot[
            style={
                color=transparent,
                draw=none,
                fill={rgb,555:red,209;green,216;blue,0},
                ,
                mark=none,
                ,
                ,
            },
            error bars/.cd,y dir=both,y explicit]
        coordinates {
            (BV, 73.44500085629635) +- (0, 19.77415840486681)
            (CNU, 40.95383359574651) +- (0, 13.990904856915233)
            (Cuccaro, 30.551968999725215) +- (0, 4.159582970140711)
            (QFT-Adder, 25.400845820861818) +- (0, 8.68665078656945)
            (QAOA, 30.501674541634223) +- (0, 36.125682919031085)
        };
        \addlegendentry{4~~~~};

        \addplot[
            style={
                color=transparent,
                draw=none,
                fill={rgb,255:red,0;green,205;blue,110},
                ,
                mark=none,
                ,
                ,
            },
            error bars/.cd,y dir=both,y explicit]
        coordinates {
            (BV, 73.80586784976124) +- (0, 20.307155957049844)
            (CNU, 39.374426660181946) +- (0, 15.428572537162786)
            (Cuccaro, 29.712331381746626) +- (0, 4.752863809180598)
            (QFT-Adder, 21.35920634856604) +- (0, 5.183467804440099)
            (QAOA, 31.349773587275855) +- (0, 35.592470620497785)
        };
        \addlegendentry{5~~~~};

        \addplot[
            style={
                color=transparent,
                draw=none,
                fill={rgb,555:red,0;green,205;blue,110},
                ,
                mark=none,
                ,
                ,
            },
            error bars/.cd,y dir=both,y explicit]
        coordinates {
            (BV, 75.07484813908557) +- (0, 20.566055221045907)
            (CNU, 34.79949867881928) +- (0, 19.514052110720712)
            (Cuccaro, 29.283280129948103) +- (0, 4.8663128989502615)
            (QFT-Adder, 16.986641251566805) +- (0, 5.720576589121614)
            (QAOA, 32.802405864806545) +- (0, 31.787101283544043)
        };
        \addlegendentry{8~~~~};

        \addplot[
            style={
                color=transparent,
                draw=none,
                fill={rgb,255:red,125;green,0;blue,255},
                ,
                mark=none,
                ,
                ,
            },
            error bars/.cd,y dir=both,y explicit]
        coordinates {
            (BV, 76.03045121088041) +- (0, 20.99486132270009)
            (CNU, 33.74510036463411) +- (0, 19.19799388137204)
            (Cuccaro, 30.14878679556564) +- (0, 4.901597480689722)
            (QFT-Adder, 16.10981237506519) +- (0, 5.962641257704104)
            (QAOA, 32.778578310597254) +- (0, 31.707854476190764)
        };
        \addlegendentry{13};
    \end{axis}

\end{tikzpicture}%
        \hfill\quad\quad\quad\quad%
%
%
%
%
\begin{tikzpicture}[baseline,scale=1,trim axis left,trim axis right]
\pgfplotsset{every tick label/.append style={font=\small}}
\pgfplotsset{every axis label/.append style={font=\small}}

    \begin{axis}[
        name=plot7,
        title={QFT-Adder Depth},
        xlabel={maximum interaction distance},
        ylabel={post-compilation depth},
        width={0.8*\columnwidth},
        height={0.5*\columnwidth},
        xmin=1, xmax=13, ymin=-1, ymax=4259,
        ,
        legend style={
            draw=none,
            at={(1,0.5)},
            anchor=east,
            font=\small},
        ,
        clip=false,
        axis line style={draw=none},
        tick style={draw=none},,
        clip, extra x ticks={1, 13}, legend style={at={(1.1,0.5)},anchor=west},
    ]
        \addplot[color={rgb,255:red,255;green,113;blue,0}, ]
            table[x=maximum-interaction-distance, y=66 post-compilation-depth 0, col sep=comma]
            {data/mid-vs-depth-qft-adder-depth.csv}
        ;
        \addlegendentry{66};

        \addplot[color={rgb,555:red,255;green,113;blue,0}, ]
            table[x=maximum-interaction-distance, y=58 post-compilation-depth 1, col sep=comma]
            {data/mid-vs-depth-qft-adder-depth.csv}
        ;
        \addlegendentry{58};

        \addplot[color={rgb,255:red,209;green,216;blue,0}, ]
            table[x=maximum-interaction-distance, y=50 post-compilation-depth 2, col sep=comma]
            {data/mid-vs-depth-qft-adder-depth.csv}
        ;
        \addlegendentry{50};

        \addplot[color={rgb,555:red,209;green,216;blue,0}, ]
            table[x=maximum-interaction-distance, y=42 post-compilation-depth 3, col sep=comma]
            {data/mid-vs-depth-qft-adder-depth.csv}
        ;
        \addlegendentry{42};

        \addplot[color={rgb,255:red,0;green,205;blue,110}, ]
            table[x=maximum-interaction-distance, y=34 post-compilation-depth 4, col sep=comma]
            {data/mid-vs-depth-qft-adder-depth.csv}
        ;
        \addlegendentry{34};

        \addplot[color={rgb,555:red,0;green,205;blue,110}, ]
            table[x=maximum-interaction-distance, y=26 post-compilation-depth 5, col sep=comma]
            {data/mid-vs-depth-qft-adder-depth.csv}
        ;
        \addlegendentry{26};

        \addplot[color={rgb,255:red,125;green,0;blue,255}, ]
            table[x=maximum-interaction-distance, y=18 post-compilation-depth 6, col sep=comma]
            {data/mid-vs-depth-qft-adder-depth.csv}
        ;
        \addlegendentry{18};

        \addplot[color={rgb,555:red,125;green,0;blue,255}, ]
            table[x=maximum-interaction-distance, y=10 post-compilation-depth 7, col sep=comma]
            {data/mid-vs-depth-qft-adder-depth.csv}
        ;
        \addlegendentry{10};

    \end{axis}

\end{tikzpicture}
        \quad\quad\quad\quad\hfill%
    }}
    \caption{Post-compilation depth across all benchmarks. On the left, the reduction in depth over the distance 1 baseline. Each bar is the average over all benchmark sizes. On the right we see a similar drop off in post-compilation depth for the QFT-Adder. We've chosen this specific benchmark to highlight the effect of restriction zones. Here we show a subset of all sizes run. Depth initially drops but for larger interaction distances some of this benefit is lost. We expect this to be more dramatic for even larger programs.}
    \label{fig:mid-v-depth}
\end{figure*}
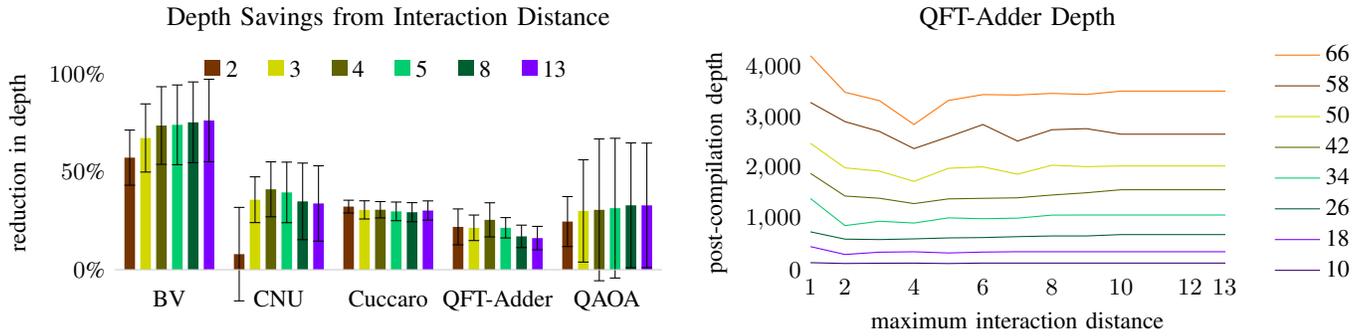

\begin{figure*}[h]
    \centering
    \scalebox{\plotscale}{%
    \makebox[1\textwidth][c]{%
        \quad\quad\quad\quad%
%
%
%
%
\begin{tikzpicture}[baseline,scale=1,trim axis left,trim axis right]
\pgfplotsset{every tick label/.append style={font=\small}}
\pgfplotsset{every axis label/.append style={font=\small}}

    \begin{axis}[
        name=plot0,
        title={Depth Increase due to Gate Serialization},
        xlabel={},
        ylabel={increase in depth},
        symbolic x coords={BV,CNU,Cuccaro,QFT-Adder,QAOA},
        width={\columnwidth},
        height={0.5*\columnwidth},
        ybar={2pt},
        bar width={4pt},
        enlargelimits=0.125,
        ymin=0, ymax=350,
        xtick=data,
        ,
        legend style={draw=none, fill=none, at={(0.5,1.03)},anchor=north,font=\small},
        legend columns=-1,
        legend image code/.code={\draw[#1, draw=none] (0em,-0.2em) rectangle (0.6em,0.4em);},
        axis line style={draw=black!20!white},
        axis on top,
        y axis line style={draw=none},
        axis x line*=bottom,
        tick style={draw=none},
        yticklabel={\pgfmathparse{\tick*1}\pgfmathprintnumber{\pgfmathresult}\%},
        clip=false,
        enlarge y limits=0,
        ,
        x tick label style={},
        grid=none,
        ymajorgrids=false,
        ,
        ,
        nodes near coords always on top/.style={
            every node near coord/.append style={
                anchor=south,
                rotate=0,
                font=\small,
                inner sep=0.2em,
            },
        },
        nodes near coords always on top,
    ]
        \addplot[
            style={
                color=transparent,
                draw=none,
                fill={rgb,555:red,255;green,113;blue,0},
                ,
                mark=none,
                ,
                ,
            },
            error bars/.cd,y dir=both,y explicit]
        coordinates {
            (BV, 9.856703197260845) +- (0, 6.635362413831602)
            (CNU, 19.722543441082557) +- (0, 4.086693502884041)
            (Cuccaro, 3.448321863758957) +- (0, 1.0509059036387294)
            (QFT-Adder, 22.183269347590993) +- (0, 9.659981808031082)
            (QAOA, 17.374680114406143) +- (0, 11.654211833486608)
        };
        \addlegendentry{2~~~~};

        \addplot[
            style={
                color=transparent,
                draw=none,
                fill={rgb,255:red,209;green,216;blue,0},
                ,
                mark=none,
                ,
                ,
            },
            error bars/.cd,y dir=both,y explicit]
        coordinates {
            (BV, 8.116638410144976) +- (0, 6.432878975547575)
            (CNU, 46.786881699312225) +- (0, 9.407192132581828)
            (Cuccaro, 8.324455053355292) +- (0, 1.2531859840996082)
            (QFT-Adder, 80.31812015193627) +- (0, 14.432730009153627)
            (QAOA, 53.52993669871307) +- (0, 27.574370141755665)
        };
        \addlegendentry{3~~~~};

        \addplot[
            style={
                color=transparent,
                draw=none,
                fill={rgb,555:red,209;green,216;blue,0},
                ,
                mark=none,
                ,
                ,
            },
            error bars/.cd,y dir=both,y explicit]
        coordinates {
            (BV, 6.141923346323844) +- (0, 6.257330526545413)
            (CNU, 68.26263199270078) +- (0, 14.616679697918656)
            (Cuccaro, 8.446480567053046) +- (0, 0.6138096964441555)
            (QFT-Adder, 130.60369514312777) +- (0, 31.138882928369583)
            (QAOA, 67.44795290339843) +- (0, 32.12451737875795)
        };
        \addlegendentry{4~~~~};

        \addplot[
            style={
                color=transparent,
                draw=none,
                fill={rgb,255:red,0;green,205;blue,110},
                ,
                mark=none,
                ,
                ,
            },
            error bars/.cd,y dir=both,y explicit]
        coordinates {
            (BV, 7.750185557742015) +- (0, 9.944646942580246)
            (CNU, 90.79019545418144) +- (0, 26.012024891764376)
            (Cuccaro, 9.86422932916861) +- (0, 1.1358835243635184)
            (QFT-Adder, 199.55691794514902) +- (0, 82.62406819085344)
            (QAOA, 95.12024269167128) +- (0, 58.6834660822649)
        };
        \addlegendentry{5~~~~};

        \addplot[
            style={
                color=transparent,
                draw=none,
                fill={rgb,555:red,0;green,205;blue,110},
                ,
                mark=none,
                ,
                ,
            },
            error bars/.cd,y dir=both,y explicit]
        coordinates {
            (BV, 4.918077072502506) +- (0, 11.50183488685184)
            (CNU, 109.30220043543517) +- (0, 32.40611942693075)
            (Cuccaro, 10.536062829111502) +- (0, 1.2167591468741978)
            (QFT-Adder, 252.50439216171202) +- (0, 125.65134464318646)
            (QAOA, 102.05019305019306) +- (0, 71.1829067709748)
        };
        \addlegendentry{8~~~~};

        \addplot[
            style={
                color=transparent,
                draw=none,
                fill={rgb,255:red,125;green,0;blue,255},
                ,
                mark=none,
                ,
                ,
            },
            error bars/.cd,y dir=both,y explicit]
        coordinates {
            (BV, 0.0) +- (0, 0.0)
            (CNU, 113.66955767712393) +- (0, 34.177413914058576)
            (Cuccaro, 9.176228566930028) +- (0, 1.42756950472355)
            (QFT-Adder, 255.74645447123407) +- (0, 124.5970476132637)
            (QAOA, 102.89253402046579) +- (0, 73.37555499659837)
        };
        \addlegendentry{13};
    \end{axis}

\end{tikzpicture}%
        \hfill\quad\quad\quad\quad%
%
%
%
%
\begin{tikzpicture}[baseline,scale=1,trim axis left,trim axis right]
\pgfplotsset{every tick label/.append style={font=\small}}
\pgfplotsset{every axis label/.append style={font=\small}}

    \begin{axis}[
        name=plot7,
        title={QAOA Depth},
        xlabel={maximum interaction distance},
        ylabel={post-compilation depth},
        width={0.8*\columnwidth},
        height={0.5*\columnwidth},
        xmin=1, xmax=13, ymin=-1, ymax=550,
        ,
        legend style={
            draw=none,
            at={(1,0.5)},
            anchor=east,
            font=\small},
        ,
        clip=false,
        axis line style={draw=none},
        tick style={draw=none},,
        clip, extra x ticks={1, 13}, legend style={at={(1.1,0.5)},anchor=west},
    ]
        \addplot[color={rgb,555:red,255;green,113;blue,0}, ]
            table[x=maximum-interaction-distance, y=50 post-compilation-depth 0, col sep=comma]
            {data/mid-vs-best-qaoa-depth.csv}
        ;
        \addlegendentry{50};

        \addplot[color={rgb,555:red,255;green,113;blue,0}, dash pattern=on 4pt off 1.6pt, ]
            table[x=maximum-interaction-distance, y=50 post-compilation-depth 1, col sep=comma]
            {data/mid-vs-best-qaoa-depth.csv}
        ;
        \addlegendentry{50};

        \addplot[color={rgb,255:red,0;green,205;blue,110}, ]
            table[x=maximum-interaction-distance, y=40 post-compilation-depth 2, col sep=comma]
            {data/mid-vs-best-qaoa-depth.csv}
        ;
        \addlegendentry{40};

        \addplot[color={rgb,255:red,0;green,205;blue,110}, dash pattern=on 4pt off 1.6pt, ]
            table[x=maximum-interaction-distance, y=40 post-compilation-depth 3, col sep=comma]
            {data/mid-vs-best-qaoa-depth.csv}
        ;
        \addlegendentry{40};

        \addplot[color={rgb,555:red,125;green,0;blue,255}, ]
            table[x=maximum-interaction-distance, y=30 post-compilation-depth 4, col sep=comma]
            {data/mid-vs-best-qaoa-depth.csv}
        ;
        \addlegendentry{30};

        \addplot[color={rgb,555:red,125;green,0;blue,255}, dash pattern=on 4pt off 1.6pt, ]
            table[x=maximum-interaction-distance, y=30 post-compilation-depth 5, col sep=comma]
            {data/mid-vs-best-qaoa-depth.csv}
        ;
        \addlegendentry{30};

        \addplot[color={rgb,255:red,255;green,113;blue,0}, ]
            table[x=maximum-interaction-distance, y=20 post-compilation-depth 6, col sep=comma]
            {data/mid-vs-best-qaoa-depth.csv}
        ;
        \addlegendentry{20};

        \addplot[color={rgb,255:red,255;green,113;blue,0}, dash pattern=on 4pt off 1.6pt, ]
            table[x=maximum-interaction-distance, y=20 post-compilation-depth 7, col sep=comma]
            {data/mid-vs-best-qaoa-depth.csv}
        ;
        \addlegendentry{20};

    \end{axis}

\end{tikzpicture}
        \quad\quad\quad\quad\hfill%
    }}
    \caption{The induced restriction zone from interaction distance increases serialization. In the prior results this is hard to discern because compared to low interaction distance the amount of gate savings translates to depth reduction. Here we compare benchmarks compiled with our restriction zone and compare to a program with no restriction zone, to mimic an ideal, highly parallel execution. The existence of a restriction zone most effect on programs which are parallel to begin with. On the right we directly compare this effect on the QAOA benchmark; solid line is compiled with realistic restriction zone and dashed is ideal. The separation between the corresponding lines signifies the effect of the restriction zone.}
    \label{fig:mid-v-best}
\end{figure*}
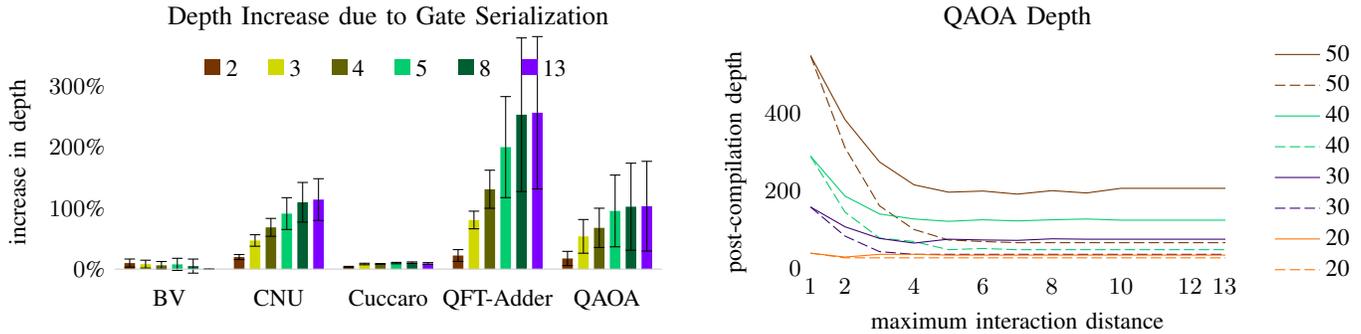

A similar trend exists for circuit depth as seen in Figure \ref{fig:mid-v-depth}. As interaction distance increases the depth tends to decrease, with the most benefit found in larger programs. Again, the rate of benefit declines quickly. We expected that as the interaction distance increased, the depth would decrease initially, then increase again due to restriction zones proportional to the interaction distance. As the maximum allowed distance increases, the average size of these zones will increase, limiting parallelism. However, there are several important factors diminishing the presence of this effect. First, SWAPs are a dominant cost in \textit{both gate count and depth}, often occurring on the critical path. Therefore, reducing the need for communication typically corresponds to a decrease in depth. Second, many quantum programs are not especially parallel and often do not contain many other gates which need to be executed at the same time limiting the potential for conflicting restriction zones. In our set of benchmarks, the circuits with high initial parallelism like CNU and QFT-Adder (a long stretch of parallel gates in the middle) do show increases in depth with increased interaction size but are not especially dramatic. In cases where gate error is dominant over coherence times, the reduction in gate count far outweighs the induced cost of greater depth or run time. 

This isn't to say there is no cost from the presence of a restriction zone. In Figure \ref{fig:mid-v-best} we analyze the relative cost of the restriction zones. In this set of experiments the program is compiled with the same maximum interaction distance. In the ideal case it is compiled with no restriction zones, resembling other architectures which permit simultaneous interactions on any set of mutually disjoint pairs. These two circuits have the same number of gates, including SWAPs. When no parallelism is lost, either from the original circuit or from parallelized communication, these lines are close. A large gap indicates the increased interaction distance causes serialization of gates. One additional side effect, which we do not model here due to complexity of simulation is the effect of crosstalk. By limiting which qubits can interact in parallel we can effectively minimize the effects of crosstalk implicitly. This can be made more explicit by artificially extending the restriction zone to reduce crosstalk error by increasing serialization. 

\subsection{Native Multiquibit Gates}
\begin{figure*}[h]
    \centering
    \scalebox{\plottightscale}{%
    \makebox[1\textwidth][c]{%
        \quad\quad\quad\quad%
%
%
%
%
\begin{tikzpicture}[baseline,scale=1,trim axis left,trim axis right]
\pgfplotsset{every tick label/.append style={font=\small}}
\pgfplotsset{every axis label/.append style={font=\small}}

    \begin{axis}[
        name=plot5,
        title={CNU Gate Count},
        xlabel={maximum interaction distance},
        ylabel={post-compilation gate count},
        width={0.25*\linewidth},
        height={0.5*\columnwidth},
        xmin=1, xmax=13, ymin=-1, ymax=2567,
        ,
        legend style={
            draw=none,
            at={(1,0.5)},
            anchor=east,
            font=\small},
        ,
        clip=false,
        axis line style={draw=none},
        tick style={draw=none},,
        clip, extra x ticks={1, 13}, legend style={at={(1,0.5)},anchor=west},
    ]
        \addplot[color={rgb,555:red,255;green,113;blue,0}, dash pattern=on 4pt off 1.6pt, ]
            table[x=maximum-interaction-distance 0, y=91 post-compilation-gate-count 0, col sep=comma]
            {data/mid-vs-3qubit-gc-cnu-gate-count.csv}
        ;
        \addlegendentry{91};

        \addplot[color={rgb,555:red,255;green,113;blue,0}, ]
            table[x=maximum-interaction-distance 1, y=91 post-compilation-gate-count 1, col sep=comma]
            {data/mid-vs-3qubit-gc-cnu-gate-count.csv}
        ;
        \addlegendentry{91};

        \addplot[color={rgb,255:red,0;green,205;blue,110}, dash pattern=on 4pt off 1.6pt, ]
            table[x=maximum-interaction-distance 2, y=59 post-compilation-gate-count 2, col sep=comma]
            {data/mid-vs-3qubit-gc-cnu-gate-count.csv}
        ;
        \addlegendentry{59};

        \addplot[color={rgb,255:red,0;green,205;blue,110}, ]
            table[x=maximum-interaction-distance 3, y=59 post-compilation-gate-count 3, col sep=comma]
            {data/mid-vs-3qubit-gc-cnu-gate-count.csv}
        ;
        \addlegendentry{59};

        \addplot[color={rgb,555:red,125;green,0;blue,255}, dash pattern=on 4pt off 1.6pt, ]
            table[x=maximum-interaction-distance 4, y=19 post-compilation-gate-count 4, col sep=comma]
            {data/mid-vs-3qubit-gc-cnu-gate-count.csv}
        ;
        \addlegendentry{19};

        \addplot[color={rgb,555:red,125;green,0;blue,255}, ]
            table[x=maximum-interaction-distance 5, y=19 post-compilation-gate-count 5, col sep=comma]
            {data/mid-vs-3qubit-gc-cnu-gate-count.csv}
        ;
        \addlegendentry{19};

    \end{axis}

\end{tikzpicture}%
        \hfill\quad\quad\quad\quad\quad\quad%
%
%
%
%
\begin{tikzpicture}[baseline,scale=1,trim axis left,trim axis right]
\pgfplotsset{every tick label/.append style={font=\small}}
\pgfplotsset{every axis label/.append style={font=\small}}

    \begin{axis}[
        name=plot5,
        title={Cuccaro Gate Count},
        xlabel={maximum interaction distance},
        ylabel={},
        width={0.25*\linewidth},
        height={0.5*\columnwidth},
        xmin=1, xmax=13, ymin=-1, ymax=3109,
        ,
        legend style={
            draw=none,
            at={(1,0.5)},
            anchor=east,
            font=\small},
        ,
        clip=false,
        axis line style={draw=none},
        tick style={draw=none},,
        clip, extra x ticks={1, 13}, legend style={at={(1,0.5)},anchor=west},
    ]
        \addplot[color={rgb,555:red,255;green,113;blue,0}, dash pattern=on 4pt off 1.6pt, ]
            table[x=maximum-interaction-distance 0, y=94  0, col sep=comma]
            {data/mid-vs-3qubit-gc-cuccaro-gate-count.csv}
        ;
        \addlegendentry{94};

        \addplot[color={rgb,555:red,255;green,113;blue,0}, ]
            table[x=maximum-interaction-distance 1, y=94  1, col sep=comma]
            {data/mid-vs-3qubit-gc-cuccaro-gate-count.csv}
        ;
        \addlegendentry{94};

        \addplot[color={rgb,255:red,0;green,205;blue,110}, dash pattern=on 4pt off 1.6pt, ]
            table[x=maximum-interaction-distance 2, y=54  2, col sep=comma]
            {data/mid-vs-3qubit-gc-cuccaro-gate-count.csv}
        ;
        \addlegendentry{54};

        \addplot[color={rgb,255:red,0;green,205;blue,110}, ]
            table[x=maximum-interaction-distance 3, y=54  3, col sep=comma]
            {data/mid-vs-3qubit-gc-cuccaro-gate-count.csv}
        ;
        \addlegendentry{54};

        \addplot[color={rgb,555:red,125;green,0;blue,255}, dash pattern=on 4pt off 1.6pt, ]
            table[x=maximum-interaction-distance 4, y=14  4, col sep=comma]
            {data/mid-vs-3qubit-gc-cuccaro-gate-count.csv}
        ;
        \addlegendentry{14};

        \addplot[color={rgb,555:red,125;green,0;blue,255}, ]
            table[x=maximum-interaction-distance 5, y=14  5, col sep=comma]
            {data/mid-vs-3qubit-gc-cuccaro-gate-count.csv}
        ;
        \addlegendentry{14};

    \end{axis}

\end{tikzpicture}%
        \hfill\quad\quad\quad\quad\quad\quad\quad\quad%
%
%
%
%
\begin{tikzpicture}[baseline,scale=1,trim axis left,trim axis right]
\pgfplotsset{every tick label/.append style={font=\small}}
\pgfplotsset{every axis label/.append style={font=\small}}

    \begin{axis}[
        name=plot5,
        title={CNU Depth},
        xlabel={maximum interaction distance},
        ylabel={post-compilation depth},
        width={0.25*\linewidth},
        height={0.5*\columnwidth},
        xmin=1, xmax=13, ymin=-1, ymax=463,
        ,
        legend style={
            draw=none,
            at={(1,0.5)},
            anchor=east,
            font=\small},
        ,
        clip=false,
        axis line style={draw=none},
        tick style={draw=none},,
        clip, extra x ticks={1, 13}, legend style={at={(1,0.5)},anchor=west},
    ]
        \addplot[color={rgb,555:red,255;green,113;blue,0}, dashed, ]
            table[x=maximum-interaction-distance 0, y=91 post-compilation-depth 0, col sep=comma]
            {data/mid-vs-3qubit-depth-cnu-depth.csv}
        ;
        \addlegendentry{91};

        \addplot[color={rgb,555:red,255;green,113;blue,0}, ]
            table[x=maximum-interaction-distance 1, y=91 post-compilation-depth 1, col sep=comma]
            {data/mid-vs-3qubit-depth-cnu-depth.csv}
        ;
        \addlegendentry{91};

        \addplot[color={rgb,255:red,0;green,205;blue,110}, dashed, ]
            table[x=maximum-interaction-distance 2, y=59 post-compilation-depth 2, col sep=comma]
            {data/mid-vs-3qubit-depth-cnu-depth.csv}
        ;
        \addlegendentry{59};

        \addplot[color={rgb,255:red,0;green,205;blue,110}, ]
            table[x=maximum-interaction-distance 3, y=59 post-compilation-depth 3, col sep=comma]
            {data/mid-vs-3qubit-depth-cnu-depth.csv}
        ;
        \addlegendentry{59};

        \addplot[color={rgb,555:red,125;green,0;blue,255}, dashed, ]
            table[x=maximum-interaction-distance 4, y=19 post-compilation-depth 4, col sep=comma]
            {data/mid-vs-3qubit-depth-cnu-depth.csv}
        ;
        \addlegendentry{19};

        \addplot[color={rgb,555:red,125;green,0;blue,255}, ]
            table[x=maximum-interaction-distance 5, y=19 post-compilation-depth 5, col sep=comma]
            {data/mid-vs-3qubit-depth-cnu-depth.csv}
        ;
        \addlegendentry{19};

    \end{axis}

\end{tikzpicture}%
        \hfill\quad\quad\quad\quad\quad\quad%
%
%
%
%
\begin{tikzpicture}[baseline,scale=1,trim axis left,trim axis right]
\pgfplotsset{every tick label/.append style={font=\small}}
\pgfplotsset{every axis label/.append style={font=\small}}

    \begin{axis}[
        name=plot5,
        title={Cuccaro Depth},
        xlabel={maximum interaction distance},
        ylabel={},
        width={0.25*\linewidth},
        height={0.5*\columnwidth},
        xmin=1, xmax=13, ymin=-1, ymax=1739,
        ,
        legend style={
            draw=none,
            at={(1,0.5)},
            anchor=east,
            font=\small},
        ,
        clip=false,
        axis line style={draw=none},
        tick style={draw=none},,
        clip, extra x ticks={1, 13}, legend style={at={(1,0.5)},anchor=west},
    ]
        \addplot[color={rgb,555:red,255;green,113;blue,0}, dashed, ]
            table[x=maximum-interaction-distance 0, y=94  0, col sep=comma]
            {data/mid-vs-3qubit-depth-cuccaro-depth.csv}
        ;
        \addlegendentry{94};

        \addplot[color={rgb,555:red,255;green,113;blue,0}, ]
            table[x=maximum-interaction-distance 1, y=94  1, col sep=comma]
            {data/mid-vs-3qubit-depth-cuccaro-depth.csv}
        ;
        \addlegendentry{94};

        \addplot[color={rgb,255:red,0;green,205;blue,110}, dashed, ]
            table[x=maximum-interaction-distance 2, y=54  2, col sep=comma]
            {data/mid-vs-3qubit-depth-cuccaro-depth.csv}
        ;
        \addlegendentry{54};

        \addplot[color={rgb,255:red,0;green,205;blue,110}, ]
            table[x=maximum-interaction-distance 3, y=54  3, col sep=comma]
            {data/mid-vs-3qubit-depth-cuccaro-depth.csv}
        ;
        \addlegendentry{54};

        \addplot[color={rgb,555:red,125;green,0;blue,255}, dashed, ]
            table[x=maximum-interaction-distance 4, y=14  4, col sep=comma]
            {data/mid-vs-3qubit-depth-cuccaro-depth.csv}
        ;
        \addlegendentry{14};

        \addplot[color={rgb,555:red,125;green,0;blue,255}, ]
            table[x=maximum-interaction-distance 5, y=14  5, col sep=comma]
            {data/mid-vs-3qubit-depth-cuccaro-depth.csv}
        ;
        \addlegendentry{14};

    \end{axis}

\end{tikzpicture}%
        \quad\quad\quad\quad%
    }}
    \caption{Compiling to programs directly to three qubit gates reduces both gate count and depth. Here we highlight a serial and parallel application written to three qubit gates. Here dashed lines are compiled to two qubit gates decomposing all Toffoli gates before mapping and routing. Solid lines compile with native Toffoli gates. With native implementation of three qubit gates we obtain huge reductions in both depth and gate count for both benchmarks.}
    \label{fig:three-comp}
\end{figure*}
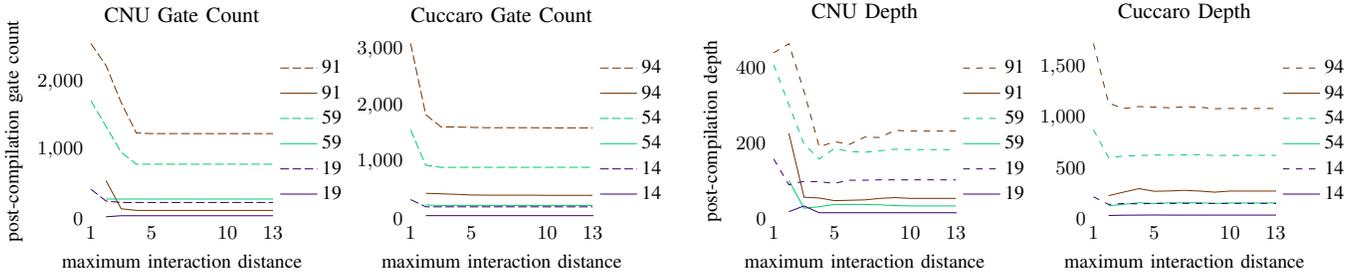
Long range interactions are not the only unique property of NA architectures. One of the important promises of NA hardware is the ability to interact multiple qubits and execute complex instructions natively. For example, gates like the three qubit Toffoli could be executed in a single step. This is important for several reasons. 

First, it doesn't require expensive decompositions to one- and two-qubit gates. Gates like the generalized Toffoli have expensive decompositions, transforming compact complex instructions into long strings of gates before SWAPs or other communication is added. The base, 3 qubit Toffoli itself requires 6 two qubit gates and interactions between every pair of qubits. If all these Toffoli gates could be executed natively without decomposition this saves up to 6x in gate count alone. Toffoli gates are fairly common in quantum algorithms extended from classical algorithms like arithmetic since they simulate logical ANDs and ORs. If even larger gates are supported, this improvement will be even larger. 

Second, efficient decomposition of multiqubit gates often requires large numbers of extra ancilla qubits. For example, in our CNU benchmark we use the logarithmic depth decomposition which requires $O(n)$ ancilla, where $n$ is the number of controls. When complex gates are executable natively, additional qubits are typically not needed, reducing space requirements for efficient implementation of gates. 

In the NA architecture, execution of these gates does come with some constraints. For example, to execute a 3 qubit Toffoli gate, each interacting qubit needs to be less than the maximum interacting distance to every other interacting qubit. Therefore, with only an interaction distance of 1 it is impossible to execute these gates and instead they must be decomposed and for larger gates more qubits will need to be brought into proximity. While not explored explicitly in this work, larger control gates will require increasingly larger interaction distances. In general, the more qubits interacting, the larger the restriction zone, increasing serialization if the qubits are too spread out.

Our set of benchmarks contains two circuit written explicitly in terms of Toffoli gates: CNU and Cuccaro. In Figure \ref{fig:three-comp} we analyze the effect of native implementation of these gates rather than decomposition. The benefit is substantial in both cases requiring many fewer gates across all maximum interaction distances. While these gates have been demonstrated, their fidelity is much lower than the demonstrated fidelity of two qubit gates. However, a simple estimation given by the product of the gate errors in the decomposition shows the fidelity of the Toffoli gate is greater than that of the decomposition. We give a more precise analysis of this effect in the next section.

Both long range interactions and native implementation of multiqubit gates prove to be very advantageous, though the benefit is tempered by a distance-dependent area of restriction which serializes communication and computation. The importance of these effects is input dependent. Programs written without multiqubit gates cannot take advantage of native implementation. Programs which are inherently serial are less affected by large restriction zones at long interaction distances. One of the most important observations is that excessively long interaction distances are not required and most benefit is obtained in the first few increases. However, as the input program size increases for larger hardware, we expect more benefit to be gained from long interaction distances. This trend is evident here where small programs have almost no benefit from increasing distance 2 to 3 but large programs nearing the device size see much more.

\section{Error Analysis of Neutral Atom Architectures}
In the previous section we explored the effect on several key circuit parameters like gate count, depth, and parallelism. These metrics are often good indicators for the success rate of programs on near and intermediate term quantum devices where gate error is relatively high and coherence times relatively low.  In the case where gate errors and coherence times are uniform across the device and comparable between technologies, these parameters are sufficient for determining advantage of one architecture over another. However, current quantum technology is still in development with some more mature than others. For example, superconducting systems and trapped ion devices have a several year head start over recently emerging neutral atoms. 

Consequently, physical properties like gate errors and coherence times are lagging a few years behind their counterparts. It is critical however to evaluate new technologies early and often to determine practical advantages at the systems level and to understand if the unique properties offered by some new technology are able to catapult the co-designed architecture ahead of its competitors. In this section, we evaluate the predicted success rate of programs compiled to a uniform piece of hardware with currently demonstrated error rates and coherence times. It is important to note the gate fidelities and $T_1$ times used as a starting point are often measured from small systems as no large scale NA architecture has been engineered to date. The average error, or $T_1$, across the hardware may have variance, as demonstrated in other publicly available technologies, though neutral atoms promises uniformity and indistinguishability in their qubits, similar to trapped ions. 

Simulating a general quantum system incurs exponential cost with the size of the system. It is impractical to model all sources of errors during computation and simplified models are typically used to predict program success rate. Here we compute the probability a program will succeed to be the probability that no gate errors happen times the probability that no decoherence errors occur. If $p_{gate, i}$ is the probability an $i$-qubit gate succeeds and $n_i$ is the number of $i$-qubit gates then the probability no gate error occurs is given by $\prod_{i} p_{gate, i}^{n_i}$. Here we consider circuits with up to $i = 3$. For neutral atoms, we consider two different coherence times for the ground state and excited state i.e. $T_{1, g}, T_{1, e}$ and $T_{2, g}, T_{2, e}$ where the ground state coherence times are often much longer than excited state coherence times. Qubits exist in the excited state when they are participating in multiqubit interactions only. The probability coherence errors occur is given as $e^{-\Delta_g / T_{1, g} - \Delta_g / T_{2, g} - \Delta_e / T_{1, e} - \Delta_e / T_{2, e}}$ where $\Delta_g$, $\Delta_e$ are the durations spent in the ground and excited states, respectively. Often, gate fidelities already include the effects of $T_1$ and $T_2$, i.e. $p_{gate, i}$ includes coherence error. Therefore, we will consider the probability of no coherence error as  $e^{-\Delta_g / T_{1, g} - \Delta_g / T_{2, g}}$ only. These simplifications serve as an upper bound approximation on the success rate of a program. 

In this section we compare against superconducting systems, specifically, using error values available via IBM for their Rome device, accessed on 11/19/2020. While we directly compare using the same simulation techniques we want to emphasize the purpose of these figures is to indicate the value gained from decreasing gate count and reducing depth relative to other available technology and to suggest that these improvements help overcome current gate error in neutral atom technology. These are not meant to suggest neutral atoms in their current stages are superior to superconducting qubits. 

Our simulation results are across a large sweep of error rates from an order of magnitude worse to many orders of magnitude better, where error rates are expected to progress to in order to make error correction feasible. The point is to evaluate different technologies with comparable error rates, how much is saved by architectural differences rather than the current status of hardware error rates, especially when neutral atoms are years behind in development.

\begin{figure}
    \centering
    \scalebox{\plotscale}{%
    \makebox[1\columnwidth][c]{%
        \hfill\quad\quad\quad\quad%
%
%
%
%
\begin{tikzpicture}[baseline,scale=1,trim axis left,trim axis right]
\pgfplotsset{every tick label/.append style={font=\small}}
\pgfplotsset{every axis label/.append style={font=\small}}

    \begin{loglogaxis}[
        name=plot11,
        title={Success Rate Comparison},
        xlabel={two-qubit gate error},
        ylabel={sample error rate},
        width={\columnwidth},
        height={0.6*\columnwidth},
        xmin=1e-05, xmax=0.1, ymin=0.03162277660168379, ymax=1.4,
        ,
        legend style={
            draw=none,
            at={(1,0)},
            anchor=south east,
            font=\small},
        ,
        clip=false,
        axis line style={draw=none},
        tick style={draw=none},,
        clip, legend style={at={(1,0)},anchor=south east,fill=none},
    ]
        \addplot[color={rgb,255:red,209;green,216;blue,0}, mark=o, mark size=2pt, dashed, every mark/.append style={solid}, mark repeat=5,mark phase=7,forget plot, ]
            table[x=two-qubit-gate-error 0, y=sample-error-rate 0, col sep=comma]
            {data/psr-all-success-rate-comparison.csv}
        ;

        \addplot[color={rgb,255:red,209;green,216;blue,0}, mark=o, mark size=2pt, mark repeat=5,mark phase=7, ]
            table[x=two-qubit-gate-error 1, y=bv sample-error-rate 1, col sep=comma]
            {data/psr-all-success-rate-comparison.csv}
        ;
        \addlegendentry{BV};

        \addplot[color={rgb,555:red,209;green,216;blue,0}, mark=x, mark size=2pt, dashed, every mark/.append style={solid}, mark repeat=5,mark phase=7,forget plot, ]
            table[x=two-qubit-gate-error 2, y=sample-error-rate 2, col sep=comma]
            {data/psr-all-success-rate-comparison.csv}
        ;

        \addplot[color={rgb,555:red,209;green,216;blue,0}, mark=x, mark size=2pt, mark repeat=5,mark phase=7, ]
            table[x=two-qubit-gate-error 3, y=cnu sample-error-rate 3, col sep=comma]
            {data/psr-all-success-rate-comparison.csv}
        ;
        \addlegendentry{CNU};

        \addplot[color={rgb,255:red,0;green,205;blue,110}, mark=+, mark size=2pt, dashed, every mark/.append style={solid}, mark repeat=5,mark phase=7,forget plot, ]
            table[x=two-qubit-gate-error 4, y=sample-error-rate 4, col sep=comma]
            {data/psr-all-success-rate-comparison.csv}
        ;

        \addplot[color={rgb,255:red,0;green,205;blue,110}, mark=+, mark size=2pt, mark repeat=5,mark phase=7, ]
            table[x=two-qubit-gate-error 5, y=cuccaro sample-error-rate 5, col sep=comma]
            {data/psr-all-success-rate-comparison.csv}
        ;
        \addlegendentry{Cuccaro};

        \addplot[color={rgb,555:red,0;green,205;blue,110}, mark=*, mark size=2pt, dashed, every mark/.append style={solid}, mark repeat=5,mark phase=7,forget plot, ]
            table[x=two-qubit-gate-error 6, y=sample-error-rate 6, col sep=comma]
            {data/psr-all-success-rate-comparison.csv}
        ;

        \addplot[color={rgb,555:red,0;green,205;blue,110}, mark=*, mark size=2pt, mark repeat=5,mark phase=7, ]
            table[x=two-qubit-gate-error 7, y=qft-adder sample-error-rate 7, col sep=comma]
            {data/psr-all-success-rate-comparison.csv}
        ;
        \addlegendentry{QFT-Adder};

        \addplot[color={rgb,255:red,125;green,0;blue,255}, mark=triangle, mark size=2pt, dashed, every mark/.append style={solid}, mark repeat=5,mark phase=7,forget plot, ]
            table[x=two-qubit-gate-error 8, y=sample-error-rate 8, col sep=comma]
            {data/psr-all-success-rate-comparison.csv}
        ;

        \addplot[color={rgb,255:red,125;green,0;blue,255}, mark=triangle, mark size=2pt, mark repeat=5,mark phase=7, ]
            table[x=two-qubit-gate-error 9, y=qaoa sample-error-rate 9, col sep=comma]
            {data/psr-all-success-rate-comparison.csv}
        ;
        \addlegendentry{QAOA};

        \addplot[color=black, dashed, ]
            table[x=two-qubit-gate-error 10, y=vs-sc-device sample-error-rate 10, col sep=comma]
            {data/psr-all-success-rate-comparison.csv}
        ;
        \addlegendentry{vs. SC device};

        \addplot[color=black, thick, dotted, ]
            table[x=two-qubit-gate-error 11, y=current-sc sample-error-rate 11, col sep=comma]
            {data/psr-all-success-rate-comparison.csv}
        ;
        \addlegendentry{\added{Current SC}};

    \end{loglogaxis}

\end{tikzpicture}%
        \hfill%
    }}
    \caption{Program success rate as a function of two-qubit error rate. Because current NA error rates are lagging behind competitive technologies we scan over a range of two-qubit error rates for each of the benchmarks all on 50 qubit programs (49 for CNU) with max interaction distance of 3. Examining pairs of solid and dashed lines we can compare NA to SC. In the limit of very low two qubit error rate, systems can support error correction. Both SC and NA systems scale at roughly the same rate (slope of the line) but the NA system diverges from the completely random outcome at higher error, allowing us to run programs on the hardware much sooner.}
    \label{fig:physical_vs_program}
\end{figure}
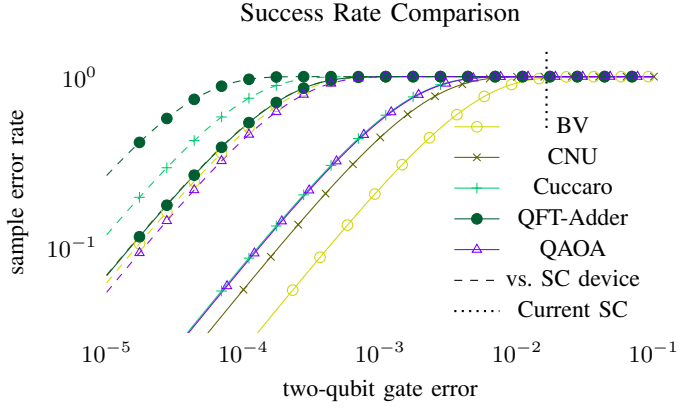

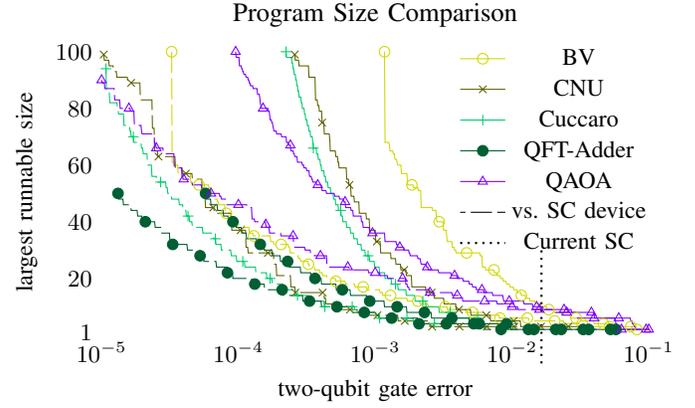
\begin{figure}
    \centering
    \scalebox{\plotscale}{%
    \makebox[1\columnwidth][c]{%
        \hfill\quad\quad\quad\quad%
%
%
%
%
\begin{tikzpicture}[baseline,scale=1,trim axis left,trim axis right]
\pgfplotsset{every tick label/.append style={font=\small}}
\pgfplotsset{every axis label/.append style={font=\small}}

    \begin{semilogxaxis}[
        name=plot11,
        title={Program Size Comparison},
        xlabel={two-qubit gate error},
        ylabel={largest runnable size},
        width={\columnwidth},
        height={0.6*\columnwidth},
        xmin=1e-05, xmax=0.1, ymin=1, ymax=100,
        ,
        legend style={
            draw=none,
            at={(1,1)},
            anchor=north east,
            font=\small},
        ,
        clip=false,
        axis line style={draw=none},
        tick style={draw=none},,
        extra y ticks={1}, legend style={at={(1.02,1.05)},anchor=north east,fill=none},
    ]
        \addplot[color={rgb,255:red,209;green,216;blue,0}, mark=o, mark size=2pt, dash pattern=on 8pt off 2pt, every mark/.append style={solid}, mark repeat=50,mark phase=0,forget plot, ]
            table[x=two-qubit-gate-error 0, y=largest-runnable-size 0, col sep=comma]
            {data/maxsize-all-program-size-comparison.csv}
        ;

        \addplot[color={rgb,255:red,209;green,216;blue,0}, mark=o, mark size=2pt, mark repeat=50,mark phase=0, ]
            table[x=two-qubit-gate-error 1, y=bv largest-runnable-size 1, col sep=comma]
            {data/maxsize-all-program-size-comparison.csv}
        ;
        \addlegendentry{BV};

        \addplot[color={rgb,555:red,209;green,216;blue,0}, mark=x, mark size=2pt, dash pattern=on 8pt off 2pt, every mark/.append style={solid}, mark repeat=50,mark phase=0,forget plot, ]
            table[x=two-qubit-gate-error 2, y=largest-runnable-size 2, col sep=comma]
            {data/maxsize-all-program-size-comparison.csv}
        ;

        \addplot[color={rgb,555:red,209;green,216;blue,0}, mark=x, mark size=2pt, mark repeat=50,mark phase=0, ]
            table[x=two-qubit-gate-error 3, y=cnu largest-runnable-size 3, col sep=comma]
            {data/maxsize-all-program-size-comparison.csv}
        ;
        \addlegendentry{CNU};

        \addplot[color={rgb,255:red,0;green,205;blue,110}, mark=+, mark size=2pt, dash pattern=on 8pt off 2pt, every mark/.append style={solid}, mark repeat=50,mark phase=0,forget plot, ]
            table[x=two-qubit-gate-error 4, y=largest-runnable-size 4, col sep=comma]
            {data/maxsize-all-program-size-comparison.csv}
        ;

        \addplot[color={rgb,255:red,0;green,205;blue,110}, mark=+, mark size=2pt, mark repeat=50,mark phase=0, ]
            table[x=two-qubit-gate-error 5, y=cuccaro largest-runnable-size 5, col sep=comma]
            {data/maxsize-all-program-size-comparison.csv}
        ;
        \addlegendentry{Cuccaro};

        \addplot[color={rgb,555:red,0;green,205;blue,110}, mark=*, mark size=2pt, dash pattern=on 8pt off 2pt, every mark/.append style={solid}, mark repeat=50,mark phase=0,forget plot, ]
            table[x=two-qubit-gate-error 6, y=largest-runnable-size 6, col sep=comma]
            {data/maxsize-all-program-size-comparison.csv}
        ;

        \addplot[color={rgb,555:red,0;green,205;blue,110}, mark=*, mark size=2pt, mark repeat=50,mark phase=0, ]
            table[x=two-qubit-gate-error 7, y=qft-adder largest-runnable-size 7, col sep=comma]
            {data/maxsize-all-program-size-comparison.csv}
        ;
        \addlegendentry{QFT-Adder};

        \addplot[color={rgb,255:red,125;green,0;blue,255}, mark=triangle, mark size=2pt, dash pattern=on 8pt off 2pt, every mark/.append style={solid}, mark repeat=50,mark phase=0,forget plot, ]
            table[x=two-qubit-gate-error 8, y=largest-runnable-size 8, col sep=comma]
            {data/maxsize-all-program-size-comparison.csv}
        ;

        \addplot[color={rgb,255:red,125;green,0;blue,255}, mark=triangle, mark size=2pt, mark repeat=50,mark phase=0, ]
            table[x=two-qubit-gate-error 9, y=qaoa largest-runnable-size 9, col sep=comma]
            {data/maxsize-all-program-size-comparison.csv}
        ;
        \addlegendentry{QAOA};

        \addplot[color=black, dash pattern=on 2pt off 2pt on 8pt off 2pt on 8pt, ]
            table[x=two-qubit-gate-error 10, y=vs-sc-device largest-runnable-size 10, col sep=comma]
            {data/maxsize-all-program-size-comparison.csv}
        ;
        \addlegendentry{vs. SC device};

        \addplot[color=black, thick, dotted, ]
            table[x=two-qubit-gate-error 11, y=current-sc largest-runnable-size 11, col sep=comma]
            {data/maxsize-all-program-size-comparison.csv}
        ;
        \addlegendentry{\added{Current SC}};

    \end{semilogxaxis}

\end{tikzpicture}%
        \hfill%
    }}
    \caption{Another way to examine the data of Figure \ref{fig:physical_vs_program} is to ask, given a desired program success rate, what the required two qubit error rate is. Here we sweep again over two qubit error rates and record the maximum program size to run with success probability greater than 2/3. Again, examining pairs of solid and dashed lines we can compare NA to SC. With the reduced gate counts and depth we expect to be able to run larger programs sooner.}
    \label{fig:max_size_ex}
\end{figure}

In Figure \ref{fig:physical_vs_program} we analyze the potential of NA architectures on three representative benchmarks. Here we sweep across various physical error rates and extract the predicted error rate with those parameters; lower is better.  In both CNU and Cuccaro we permit 3 qubit gates while the others contain only one- and two- qubit gates. The superconducting curves correspond to similar simulations using error rates and coherence times provided by IBM. At lower physical error rates we expect all architectures to perform well, at virtually no program error rate. At this limit, the hardware is below the threshold for error correction. On the other hand, in the limit of high physical error rates, we expect no program to succeed with any likelihood and to produce random results. For near- and intermediate-term quantum computation, the regions between the limits are most important and the divergence from this all-noise outcome determines how quickly a device becomes viable for practical computation. For comparable error rates between superconducting and NA architectures, we see great advantage obtained via long range interactions and native multiqubit gates, diverging more quickly than the limited connectivity SC devices.

Alternatively, we might ask what physical error rates are needed to run programs of a given size with probability of success above some threshold. In Figure \ref{fig:max_size_ex}, we consider this question for a threshold success rate of $2/3$. Here we sweep across physical error rates and compute the largest possible program of each benchmark we can successfully execute. There are two interpretations. First, for a fixed physical error rate we can determine what size program is likely to be successfully executed. Alternatively, suppose we have a program of a given size we want to execute, we can then decide what physical error rate do we need to run that program successfully. For a fixed error rate, we find we can execute a larger program or, equivalently, require worse physical error rates than a superconducting system to run a desired program.

\begin{figure*}
    \centering
    \scalebox{1}{
        \def\svgwidth{1.3\columnwidth}
        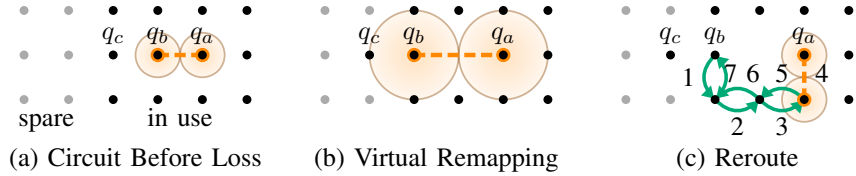%
    }
    \caption{Examples of two different atom loss coping strategies. (a) shows the initial configuration of three qubits, with the spare qubits in a light grey, and in use qubits black. (b) Represents how the atoms are shifted into the spare qubits to accommodate a lost atom under the virtual remapping strategy. Notice that the interaction is no longer within interaction distance 1. (c) Demonstrates how the qubits can be swapped to a valid interaction configuration, and returned for rerouting strategies.  Numbers indicate the order of swaps.}
    \label{fig:loss-strategies}
\end{figure*}

\section{Unique Challenge: Sporadic Atom Loss}
So far, we've focused primarily on properties of a neutral atom system that are usually advantageous and have analyzed that while there are tradeoffs, the gates and depth saved drastically outweighs any cost. However, neutral atom systems are not without limitation. The atoms in a NA system are trapped using optical dipole traps such as optical tweezers. While this technique offers great flexibility in array geometries and the prospect of scaling to large atom counts, the trapping potential per atom is weak when compared to trapped ion architectures. Consequently, neutral atoms can be lost more easily during or between computations forming a sparser grid. Fortunately, this loss can be detected via fluorescence imaging and remedied with various software and hardware approaches with different overhead costs analyzed here.

Atom loss has critical implications on program execution. For near-term computation, we run programs hundreds or thousands of times to obtain a distribution of answers. Consider running a program, after each trial we evaluate whether atoms have been lost. If an atom used by the computation has been lost, then we have an incomplete answer. Upon a loss, we have no way of knowing if it occurred in computation and must disregard the run from our distribution and perform another shot. Furthermore, a program compiled for the original grid of qubits may no longer be executable. The program must either be recompiled for the now sparser grid of qubits, the array of atoms can be reloaded, or the compiled circuit can be adapted to the atom loss. The first two solutions are costly in terms of overhead time.  The third may provide a faster alternative, and provide opportunity to perform more executions in the same time while maintaining a valid program.

We model atom loss from two processes. The first is based on vacuum limited lifetime where there is a finite chance a background atom collides with the qubit atom held in the optical tweezers displacing the qubit. We approximate this occurs with probability $0.0068$ over the course of a program and is uniform across all qubits \cite{vacuum-atom-loss}. Loss during readout is much more likely. In some systems readout occurs by ejecting atoms which are not in a given state, resulting in about 50\% atom loss every cycle \cite{meas-loss-1}. This model is extremely destructive and coping strategies are only effective if the program is much smaller than the total size of the hardware. Alternative, potentially lossless techniques, have been proposed for measurement, but are not perfect with loss approximately $2$\% \cite{meas-loss-2} uniformly across all measured atoms. Detecting atom loss is done via fluorescence which takes on the order of $6$ms.

We propose several coping mechanisms to the loss of atoms and examine their effectiveness in terms of overheads such as the time to perform array loads, qubit fluorescence, and potential recompilation. In each of these experiments we assume the input program is smaller than the total number of hardware qubits in the \textit{original} grid, otherwise any loss of an atom \textit{requires} a reload. These additional unused qubits after initial compilation are considered \textit{spares}, borrowing from classical work on DRAM sparing \cite{dram-spare}. Currently, no program that executes with reasonably high success will use the entire grid and these spares will come at no cost. Potentially, as many atom losses can be sustained as number of spares. However, an array reload is always possible there is an unlucky set of holes or no more spares. Below we detail various strategies.

\begin{figure*}[h]
    \centering
    \scalebox{\plotscale}{%
    \makebox[1\textwidth][c]{%
        \quad\quad\quad\quad%
%
%
%
%
\begin{tikzpicture}[baseline,scale=1,trim axis left,trim axis right]
\pgfplotsset{every tick label/.append style={font=\small}}
\pgfplotsset{every axis label/.append style={font=\small}}

    \begin{axis}[
        name=plot0,
        title={Max Atom Loss Tolerance (CNU)},
        xlabel={maximum interaction distance},
        ylabel={\# atoms lost/device size},
        symbolic x coords={2,3,4,5,6},
        width={\columnwidth},
        height={0.5*\columnwidth},
        ybar={2pt},
        bar width={4pt},
        enlargelimits=0.125,
        ymin=0, ymax=80,
        xtick=data,
        ,
        legend style={draw=none, fill=none, at={(0.5,1.03)},anchor=north,font=\small},
        legend columns=-1,
        legend image code/.code={\draw[#1, draw=none] (0em,-0.2em) rectangle (0.6em,0.4em);},
        axis line style={draw=black!20!white},
        axis on top,
        y axis line style={draw=none},
        axis x line*=bottom,
        tick style={draw=none},
        yticklabel={\pgfmathparse{\tick*1}\pgfmathprintnumber{\pgfmathresult}\%},
        clip=false,
        enlarge y limits=0,
        ,
        x tick label style={},
        grid=none,
        ymajorgrids=false,
        legend columns=10,legend cell align={left}, legend style={at={(1.1,1.05)},anchor=north},
        ,
        nodes near coords always on top/.style={
            every node near coord/.append style={
                anchor=south,
                rotate=0,
                font=\small,
                inner sep=0.2em,
            },
        },
        nodes near coords always on top,
    ]
        \addplot[
            style={
                color=transparent,
                draw=none,
                fill={rgb,255:red,209;green,216;blue,0},
                ,
                mark=none,
                ,
                ,
            },
            error bars/.cd,y dir=both,y explicit]
        coordinates {
            (2, 2.116) +- (0, 2.8295860323072803)
            (3, 3.528) +- (0, 4.023131513429395)
            (4, 5.986) +- (0, 6.145676747549511)
            (5, 3.352) +- (0, 4.198016517996415)
            (6, 11.182) +- (0, 10.109266567910955)
        };
        \addlegendentry{virtual remapping~~~~};

        \addplot[
            style={
                color=transparent,
                draw=none,
                fill={rgb,555:red,209;green,216;blue,0},
                ,
                mark=none,
                ,
                ,
            },
            error bars/.cd,y dir=both,y explicit]
        coordinates {
            (2, 49.266) +- (0, 7.660326703895914)
            (3, 50.678) +- (0, 8.070227577106644)
            (4, 50.026) +- (0, 8.243131834739135)
            (5, 50.2) +- (0, 8.002003757075881)
            (6, 50.164) +- (0, 7.828132812277351)
        };
        \addlegendentry{reroute~~~~};

        \addplot[
            style={
                color=transparent,
                draw=none,
                fill={rgb,255:red,0;green,205;blue,110},
                ,
                mark=none,
                ,
                ,
            },
            error bars/.cd,y dir=both,y explicit]
        coordinates {
            (2, nan) +- (0, 0.0)
            (3, 11.36) +- (0, 6.713907439502902)
            (4, 16.686) +- (0, 8.582932918003607)
            (5, 23.722) +- (0, 10.065552478298782)
            (6, 28.664) +- (0, 11.854626452852632)
        };
        \addlegendentry{compile small~~~~};

        \addplot[
            style={
                color=transparent,
                draw=none,
                fill={rgb,555:red,0;green,205;blue,110},
                ,
                mark=none,
                ,
                ,
            },
            error bars/.cd,y dir=both,y explicit]
        coordinates {
            (2, nan) +- (0, 0.0)
            (3, 51.074) +- (0, 7.736422200016867)
            (4, 50.686) +- (0, 8.078485492696723)
            (5, 50.026) +- (0, 8.243131834739135)
            (6, 50.2) +- (0, 8.002003757075881)
        };
        \addlegendentry{c. small+reroute~~~~};

        \addplot[
            style={
                color=transparent,
                draw=none,
                fill={rgb,255:red,125;green,0;blue,255},
                ,
                mark=none,
                ,
                ,
            },
            error bars/.cd,y dir=both,y explicit]
        coordinates {
            (2, 44.41) +- (0, 8.316021430037907)
            (3, 66.534) +- (0, 4.76005885684634)
            (4, 69.854) +- (0, 0.852136299276759)
            (5, 69.998) +- (0, 0.044721359549995794)
            (6, 70.0) +- (0, 0.0)
        };
        \addlegendentry{recompile};
    \end{axis}

\end{tikzpicture}%
        \hfill\quad\quad\quad\quad%
%
%
%
%
\begin{tikzpicture}[baseline,scale=1,trim axis left,trim axis right]
\pgfplotsset{every tick label/.append style={font=\small}}
\pgfplotsset{every axis label/.append style={font=\small}}

    \begin{axis}[
        name=plot0,
        title={Max Atom Loss Tolerance (Cuccaro)},
        xlabel={maximum interaction distance},
        ylabel={},
        symbolic x coords={2,3,4,5,6},
        width={\columnwidth},
        height={0.5*\columnwidth},
        ybar={2pt},
        bar width={4pt},
        enlargelimits=0.125,
        ymin=0, ymax=80,
        xtick=data,
        ,
        legend style={draw=none, fill=none, at={(0.5,1.03)},anchor=north,font=\small},
        legend columns=-1,
        legend image code/.code={\draw[#1, draw=none] (0em,-0.2em) rectangle (0.6em,0.4em);},
        axis line style={draw=black!20!white},
        axis on top,
        y axis line style={draw=none},
        axis x line*=bottom,
        tick style={draw=none},
        yticklabel={\pgfmathparse{\tick*1}\pgfmathprintnumber{\pgfmathresult}\%},
        clip=false,
        enlarge y limits=0,
        ,
        x tick label style={},
        grid=none,
        ymajorgrids=false,
        legend columns=10,legend cell align={left}, legend style={at={(1.1,1.05)},anchor=north}, yticklabels={,,},
        ,
        nodes near coords always on top/.style={
            every node near coord/.append style={
                anchor=south,
                rotate=0,
                font=\small,
                inner sep=0.2em,
            },
        },
        nodes near coords always on top,
    ]
        \addplot[
            style={
                color=transparent,
                draw=none,
                fill={rgb,255:red,209;green,216;blue,0},
                ,
                mark=none,
                ,
                ,
            },
            error bars/.cd,y dir=both,y explicit]
        coordinates {
            (2, 2.164) +- (0, 3.052825492387986)
            (3, 3.912) +- (0, 4.5029778612867375)
            (4, 4.046) +- (0, 4.736322360122672)
            (5, 15.644) +- (0, 11.52731634180689)
            (6, 37.034) +- (0, 10.35235173150263)
        };

        \addplot[
            style={
                color=transparent,
                draw=none,
                fill={rgb,555:red,209;green,216;blue,0},
                ,
                mark=none,
                ,
                ,
            },
            error bars/.cd,y dir=both,y explicit]
        coordinates {
            (2, 47.316) +- (0, 8.15654353294394)
            (3, 49.046) +- (0, 8.166349400132846)
            (4, 48.322) +- (0, 8.084865248857545)
            (5, 48.238) +- (0, 7.918760606248819)
            (6, 48.238) +- (0, 7.918760606248819)
        };

        \addplot[
            style={
                color=transparent,
                draw=none,
                fill={rgb,255:red,0;green,205;blue,110},
                ,
                mark=none,
                ,
                ,
            },
            error bars/.cd,y dir=both,y explicit]
        coordinates {
            (2, nan) +- (0, 0.0)
            (3, 13.63) +- (0, 7.331890153736136)
            (4, 20.086) +- (0, 9.2371670174907)
            (5, 22.736) +- (0, 10.816667906948906)
            (6, 38.26) +- (0, 9.968909585136846)
        };

        \addplot[
            style={
                color=transparent,
                draw=none,
                fill={rgb,555:red,0;green,205;blue,110},
                ,
                mark=none,
                ,
                ,
            },
            error bars/.cd,y dir=both,y explicit]
        coordinates {
            (2, nan) +- (0, 0.0)
            (3, 48.74) +- (0, 8.350560683418504)
            (4, 49.054) +- (0, 8.17611041774304)
            (5, 48.322) +- (0, 8.084865248857545)
            (6, 48.238) +- (0, 7.918760606248819)
        };

        \addplot[
            style={
                color=transparent,
                draw=none,
                fill={rgb,255:red,125;green,0;blue,255},
                ,
                mark=none,
                ,
                ,
            },
            error bars/.cd,y dir=both,y explicit]
        coordinates {
            (2, 44.292) +- (0, 8.413579061567704)
            (3, 66.53) +- (0, 3.9785371074089606)
            (4, 68.946) +- (0, 0.5510820109253531)
            (5, 69.0) +- (0, 0.0)
            (6, 69.0) +- (0, 0.0)
        };
    \end{axis}

\end{tikzpicture}
        \hfill%
    }}
    \caption{Atom loss as a percentage of total device size which can be sustained before a reload of the array is needed. Each program is 30 qubits on a 100 qubit device. As the interaction distance increases most strategies can sustain more atom loss. Strategies like full recompilation can sustain large numbers of atom loss but as we will see are expensive computationally. Fast, hardware solutions or hybrid solutions can sustain fewer numbers of holes but have lower overhead. We show two representative benchmarks parallel vs. serial.}
    \label{fig:sustained-holes}
\end{figure*}
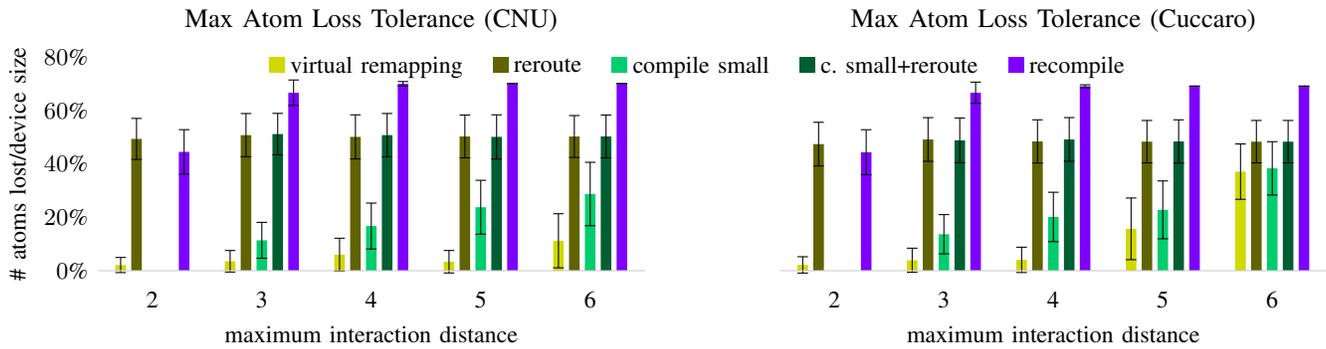

\begin{itemize}
    \item \textit{Always Reload.} Every time an atom loss is detected for a qubit used by the compiled program we reload the entire array. This naive strategy is efficient when array reloads are fast since only a single compilation step is needed.
    \item \textit{Always Full Recompile.} When an interfering atom loss is detected we update the hardware topology accordingly and recompile the input program. This fails when the topology becomes disconnected, requiring a reload. 
    \item \textit{Virtual Remapping.} NA architectures support long range interactions up to a maximum interaction distance. Therefore, shifts in the qubit placement is only detrimental to execution when the required qubit interactions exceed this distance. For this strategy, we start with a virtual mapping of physical qubits to physical qubits where each qubit maps to itself. When an atom is lost we check if it is used in the program. If so, we adjust the virtual mapping by shifting the qubits in a column or row of the qubit in the cardinal direction with the most unused qubits starting from the lost atom to the edge of the device. This process is shown in 
    Figure \ref{fig:loss-strategies}b. In this figure, addressing $q_b$ would now point to the original location of $q_c$, and addressing $q_c$ would point to the qubit to the left of $q_c$'s original location. If there are no spare qubits, we perform a full reload. Otherwise, we execute the gates in order according to the mapping. If two qubits which must interact are now too far apart, we reload. This strategy is efficient in terms of overhead since this virtual remapping can be done on the order of 40 ns in hardware \cite{dram-time} via a lookup table. However, this strategy can be inefficient in number of reloads required since it is easy to exceed the interaction distance. We later explore how many atom losses can be sustained before reload which allows us to estimate how many reloads will be required on average.
    \item \textit{Minor Rerouting.} Here we perform the same shifting strategy as in \textit{Virtual Remapping} to occupy available spares. However, rather than failing when distance between the remapped qubits exceeds the maximum interaction distance, we attempt to find a path over the usable qubits, and insert SWAP gates along this path. To simplify computation we SWAP the qubits on the found path, execute the desired gate, then reverse the process to maintain the expected mapping. The rerouting is shown in Figure \ref{fig:loss-strategies}c. Too many additional SWAPs are detrimental to program success. We may force a reload if the expected success rate drops, for example by half, from the original program's expected success rate.
    \item \textit{Compile to Smaller Than Max Interaction Distance.} For most programs there are diminishing returns to compiling to larger max interaction distances, but can be sensitive to atom loss. In this strategy we compile to an interaction distance less than the max so when qubits get shifted away from each other it will take more shifts to exceed the true maximum distance. The overhead is the same as \textit{Virtual Remapping} but the compiled program could be less efficient than one compiled to the true max distance.
    \item \textit{Compile Small and Minor Reroute}. This strategy is based on \textit{Compile to Smaller Than Max Interaction Distance} but performs the same rerouting strategy as \textit{Minor Rerouting} with similar overhead costs.
\end{itemize}

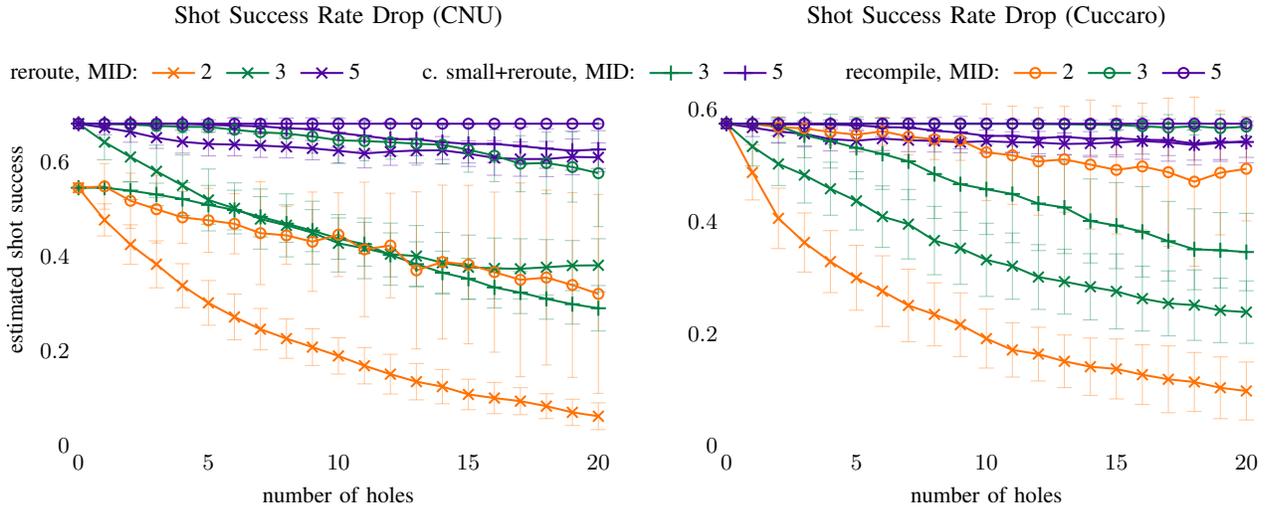
\begin{figure*}[h]
    \centering
    \scalebox{0.95}{%
    \makebox[1\textwidth][c]{%
        \quad\quad\quad\quad%
%
%
%
%
\begin{tikzpicture}[baseline,scale=1,trim axis left,trim axis right]
\pgfplotsset{every tick label/.append style={font=\small}}
\pgfplotsset{every axis label/.append style={font=\small}}

    \begin{axis}[
        name=plot10,
        title={Shot Success Rate Drop (CNU)},
        xlabel={number of holes},
        ylabel={estimated shot success},
        width={\columnwidth},
        height={0.8*\columnwidth},
        xmin=0, xmax=20, ymin=0, ymax=0.8319836611490459,
        ,
        legend style={
            draw=none,
            at={(0.5,1)},
            anchor=north,
            font=\small},
        ,
        clip=false,
        axis line style={draw=none},
        tick style={draw=none},,
        legend columns=30, legend cell align={right}, legend style={at={(1,1)},anchor=north}, mark size=3pt,
    ]
        \addplot[color=white, ]
            table[x=number-of-holes 0, y=reroute-mid estimated-shot-success 0, col sep=comma]
            {data/hole-fidelity-cnu-shot-success-rate-drop-cnu.csv}
        ;
        \addlegendentry{reroute, MID:~~};

        \addplot[color={rgb,255:red,255;green,113;blue,0}, thick, mark=x, error bars/.cd,y dir=both,y explicit, error bar style={opacity=0.4}]
            table[x=number-of-holes 1, y=2 estimated-shot-success 1, y error=2 estimated-shot-success 1-err, col sep=comma]
            {data/hole-fidelity-cnu-shot-success-rate-drop-cnu.csv}
        ;
        \addlegendentry{2~~};

        \addplot[color={rgb,405:red,0;green,205;blue,110}, thick, mark=x, error bars/.cd,y dir=both,y explicit, error bar style={opacity=0.4}]
            table[x=number-of-holes 2, y=3 estimated-shot-success 2, y error=3 estimated-shot-success 2-err, col sep=comma]
            {data/hole-fidelity-cnu-shot-success-rate-drop-cnu.csv}
        ;
        \addlegendentry{3~~};

        \addplot[color={rgb,405:red,125;green,0;blue,255}, thick, mark=x, error bars/.cd,y dir=both,y explicit, error bar style={opacity=0.4}]
            table[x=number-of-holes 3, y=5 estimated-shot-success 3, y error=5 estimated-shot-success 3-err, col sep=comma]
            {data/hole-fidelity-cnu-shot-success-rate-drop-cnu.csv}
        ;
        \addlegendentry{5~~};

        \addplot[color=white, ]
            table[x=number-of-holes 4, y=c-smallreroute-mid estimated-shot-success 4, col sep=comma]
            {data/hole-fidelity-cnu-shot-success-rate-drop-cnu.csv}
        ;
        \addlegendentry{c. small+reroute, MID:~~};

        \addplot[color={rgb,405:red,0;green,205;blue,110}, thick, mark=+, error bars/.cd,y dir=both,y explicit, error bar style={opacity=0.4}]
            table[x=number-of-holes 5, y=3 estimated-shot-success 5, y error=3 estimated-shot-success 5-err, col sep=comma]
            {data/hole-fidelity-cnu-shot-success-rate-drop-cnu.csv}
        ;
        \addlegendentry{3~~};

        \addplot[color={rgb,405:red,125;green,0;blue,255}, thick, mark=+, error bars/.cd,y dir=both,y explicit, error bar style={opacity=0.4}]
            table[x=number-of-holes 6, y=5 estimated-shot-success 6, y error=5 estimated-shot-success 6-err, col sep=comma]
            {data/hole-fidelity-cnu-shot-success-rate-drop-cnu.csv}
        ;
        \addlegendentry{5~~};

        \addplot[color=white, ]
            table[x=number-of-holes 7, y=recompile-mid estimated-shot-success 7, col sep=comma]
            {data/hole-fidelity-cnu-shot-success-rate-drop-cnu.csv}
        ;
        \addlegendentry{recompile, MID:~~};

        \addplot[color={rgb,255:red,255;green,113;blue,0}, thick, mark=o, mark size=2pt, error bars/.cd,y dir=both,y explicit, error bar style={opacity=0.4}]
            table[x=number-of-holes 8, y=2 estimated-shot-success 8, y error=2 estimated-shot-success 8-err, col sep=comma]
            {data/hole-fidelity-cnu-shot-success-rate-drop-cnu.csv}
        ;
        \addlegendentry{2~~};

        \addplot[color={rgb,405:red,0;green,205;blue,110}, thick, mark=o, mark size=2pt, error bars/.cd,y dir=both,y explicit, error bar style={opacity=0.4}]
            table[x=number-of-holes 9, y=3 estimated-shot-success 9, y error=3 estimated-shot-success 9-err, col sep=comma]
            {data/hole-fidelity-cnu-shot-success-rate-drop-cnu.csv}
        ;
        \addlegendentry{3~~};

        \addplot[color={rgb,405:red,125;green,0;blue,255}, thick, mark=o, mark size=2pt, error bars/.cd,y dir=both,y explicit, error bar style={opacity=0.4}]
            table[x=number-of-holes 10, y=5 estimated-shot-success 10, y error=5 estimated-shot-success 10-err, col sep=comma]
            {data/hole-fidelity-cnu-shot-success-rate-drop-cnu.csv}
        ;
        \addlegendentry{5~~};

    \end{axis}

\end{tikzpicture}%
        \hfill\quad\quad\quad\quad%
%
%
%
%
\begin{tikzpicture}[baseline,scale=1,trim axis left,trim axis right]
\pgfplotsset{every tick label/.append style={font=\small}}
\pgfplotsset{every axis label/.append style={font=\small}}

    \begin{axis}[
        name=plot10,
        title={Shot Success Rate Drop (Cuccaro)},
        xlabel={number of holes},
        ylabel={},
        width={\columnwidth},
        height={0.8*\columnwidth},
        xmin=0, xmax=20, ymin=0, ymax=0.7002145441499364,
        ,
        legend style={
            draw=none,
            at={(0.5,1)},
            anchor=north,
            font=\small},
        ,
        clip=false,
        axis line style={draw=none},
        tick style={draw=none},,
        legend columns=30, legend cell align={right}, legend style={at={(1,1)},anchor=north}, mark size=3pt,
    ]
        \addplot[color=white, ]
            table[x=number-of-holes 0, y=0, col sep=comma]
            {data/hole-fidelity-cuccaro-shot-success-rate-drop-cuccaro.csv}
        ;

        \addplot[color={rgb,255:red,255;green,113;blue,0}, thick, mark=x, error bars/.cd,y dir=both,y explicit, error bar style={opacity=0.4}]
            table[x=number-of-holes 1, y=1, y error=1-err, col sep=comma]
            {data/hole-fidelity-cuccaro-shot-success-rate-drop-cuccaro.csv}
        ;

        \addplot[color={rgb,405:red,0;green,205;blue,110}, thick, mark=x, error bars/.cd,y dir=both,y explicit, error bar style={opacity=0.4}]
            table[x=number-of-holes 2, y=2, y error=2-err, col sep=comma]
            {data/hole-fidelity-cuccaro-shot-success-rate-drop-cuccaro.csv}
        ;

        \addplot[color={rgb,405:red,125;green,0;blue,255}, thick, mark=x, error bars/.cd,y dir=both,y explicit, error bar style={opacity=0.4}]
            table[x=number-of-holes 3, y=3, y error=3-err, col sep=comma]
            {data/hole-fidelity-cuccaro-shot-success-rate-drop-cuccaro.csv}
        ;

        \addplot[color=white, ]
            table[x=number-of-holes 4, y=4, col sep=comma]
            {data/hole-fidelity-cuccaro-shot-success-rate-drop-cuccaro.csv}
        ;

        \addplot[color={rgb,405:red,0;green,205;blue,110}, thick, mark=+, error bars/.cd,y dir=both,y explicit, error bar style={opacity=0.4}]
            table[x=number-of-holes 5, y=5, y error=5-err, col sep=comma]
            {data/hole-fidelity-cuccaro-shot-success-rate-drop-cuccaro.csv}
        ;

        \addplot[color={rgb,405:red,125;green,0;blue,255}, thick, mark=+, error bars/.cd,y dir=both,y explicit, error bar style={opacity=0.4}]
            table[x=number-of-holes 6, y=6, y error=6-err, col sep=comma]
            {data/hole-fidelity-cuccaro-shot-success-rate-drop-cuccaro.csv}
        ;

        \addplot[color=white, ]
            table[x=number-of-holes 7, y=7, col sep=comma]
            {data/hole-fidelity-cuccaro-shot-success-rate-drop-cuccaro.csv}
        ;

        \addplot[color={rgb,255:red,255;green,113;blue,0}, thick, mark=o, mark size=2pt, error bars/.cd,y dir=both,y explicit, error bar style={opacity=0.4}]
            table[x=number-of-holes 8, y=8, y error=8-err, col sep=comma]
            {data/hole-fidelity-cuccaro-shot-success-rate-drop-cuccaro.csv}
        ;

        \addplot[color={rgb,405:red,0;green,205;blue,110}, thick, mark=o, mark size=2pt, error bars/.cd,y dir=both,y explicit, error bar style={opacity=0.4}]
            table[x=number-of-holes 9, y=9, y error=9-err, col sep=comma]
            {data/hole-fidelity-cuccaro-shot-success-rate-drop-cuccaro.csv}
        ;

        \addplot[color={rgb,405:red,125;green,0;blue,255}, thick, mark=o, mark size=2pt, error bars/.cd,y dir=both,y explicit, error bar style={opacity=0.4}]
            table[x=number-of-holes 10, y=10, y error=10-err, col sep=comma]
            {data/hole-fidelity-cuccaro-shot-success-rate-drop-cuccaro.csv}
        ;

    \end{axis}

\end{tikzpicture}
        \hfill%
    }}
    \caption{For strategies which modify the program such as recompilation or rerouting strategies, additional gates could be added leading to a lower overall success rate. Here we trace the success rate of our three program modifying strategies. The full recompilation strategy (circles) is a rough upper bound which best accounts for holes as they appear being able to move the entire program to a more appropriate location and route best. The gap between strategies on the same MID gets smaller as the MID gets larger. Here we've chosen the two-qubit error rate corresponding to approximate 0.6 success rate to begin with (based on Figure \ref{fig:max_size_ex}) in order to best demonstrate the change in shot success probability over a range of atom loss.}
    \label{fig:hole-fidelity}
\end{figure*}

Excluding the first approach of reloading with any interfering atom loss, we examine how many losses can be sustained without exceeding the constraints of the architecture in size, dimension, or needed interaction distances.  Figure \ref{fig:sustained-holes} shows the maximum number of holes supported by the different strategies for a 30 qubit Cuccaro adder and a 29 qubit CNU.  The entries for \textit{compile small} and \textit{compile small + reroute} are compiled to one less than the maximum interaction distance.  We do not compile to interaction distance 1, so we do not have entries for these strategies at interaction distance 2.

As would be expected, \textit{recompile} is able to support the most lost atoms since the only failure cases are: disconnected hardware topology, or fewer atoms than required qubits.  In fact, since our example circuits use 30\% and 29\% of the hardware, once the interaction distance overcomes any disconnected pieces, recompiling can sustain 70\% atom loss, the ideal case for sustained loss.  The non-rerouting strategies, while a fast solution, offer limited atom loss recovery.  The simple virtual remapping is only able to support a small amount of atom loss, but does increase as the max distance increases.  As predicted, compiling to a smaller interaction distance does enable more resilience to atom loss since more movement can be tolerated before exceeding the maximum interaction distance. Both rerouting strategies have a disconnected topology failure case, but also the additional failure case of not having the space in any direction to shift the qubits in the event of atom loss.  As a result, both are only able to sustain 50\% atom loss at higher interaction distances.
    
However, these different strategies add varying numbers of extra SWAPs to handle atom loss, lowering the success rate.  As more atoms are lost, more SWAPs are needed, and the rate decreases as seen in Figure \ref{fig:hole-fidelity} for Cuccaro and CNU with the rerouting and recompiling strategies.  With current error rates, success is very low for 30 qubit circuits.  To better demonstrate how the success rate changes, we use lower error rates so about $2/3$ of shots succeed without atom loss.  For any strategy, as the interaction distance increases, fewer SWAPs are needed so the shot success stays higher.  Since the recompilation strategy is able to schedule and map qubits with full knowledge of the current state, including missing atoms, it as able to achieve the best routing and success rate out of all the atom loss strategies.  Both of the rerouting strategies have lower rates since they tend to add more SWAPs per atom loss.  But, since compiling to a smaller MID before rerouting means the interaction distance is exceeded less often, it requires fewer SWAPs, boosting its rate over simply rerouting.
    
\begin{figure}
    \centering
    \scalebox{\plotscale}{%
    \makebox[1\columnwidth][c]{%
        \hfill\quad\quad\quad\quad%
        \input{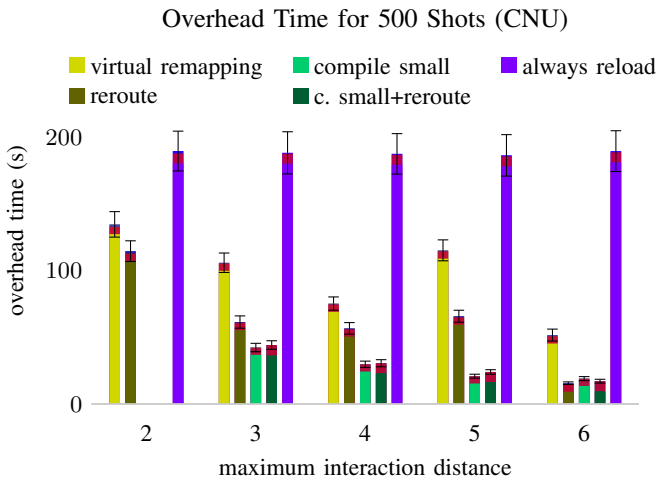}%
        \hfill%
    }}
    \caption{Strategies that are able reduce the number of reloads necessary greatly reduce the overhead time when running circuits.  Here we show the overhead time for all strategies except recompilation. The proportion of time dedicated to reloading is shown by the dominate color in each bar, followed by fluorescence in red, and recompilation in black. Any strategy whose overhead exceeds that of always reloading, such as full recompilation, should not be considered.}
    \label{fig:hole-timing}
\end{figure}

Taking this into account, we examine the estimated overhead time of each strategy for 500 runs of a given circuit.  We use a 2\% chance of atom loss for a measured qubit, and a 0.0068\% chance of atom loss due to atom collision in a vacuum.  The overhead times for CNU are seen in Figure \ref{fig:hole-timing}.  For any rerouting strategy that requires extra SWAPs, a reload is forced once the number of added swaps would decrease the success rate by 50\%.  For a 96.5\% successful two-qubit gate, this would be six SWAPs.

\begin{figure}
    \centering
    \scalebox{\plotscale}{%
    \makebox[1\columnwidth][c]{%
        \hfill\quad\quad\quad\quad%
%
%
%
%
\begin{tikzpicture}[baseline,scale=1,trim axis left,trim axis right]
\pgfplotsset{every tick label/.append style={font=\small}}
\pgfplotsset{every axis label/.append style={font=\small}}

    \begin{loglogaxis}[
        name=plot3,
        title={Successful Shot Rate Sensitivity (CNU)},
        xlabel={atom loss rate improvement factor},
        ylabel={\# successful shots before reload},
        width={\columnwidth},
        height={0.5*\columnwidth},
        ymax=200,
        ,
        legend style={
            draw=none,
            at={(1,0)},
            anchor=south east,
            font=\small},
        ,
        clip=false,
        axis line style={draw=none},
        tick style={draw=none},,
        ymajorgrids=true, extra y ticks={1, 0.01, 100},
    ]
        \addplot[color={rgb,255:red,255;green,113;blue,0}, error bars/.cd,y dir=both,y explicit]
            table[x=atom-loss-rate-improvement-factor, y=6 -successful-shots-before-reload 0, y error=6 -successful-shots-before-reload 0-err, col sep=comma]
            {data/hole-sensitivity-cnu-successful-shot-rate-sensitivity-cnu.csv}
        ;
        \addlegendentry{6};

        \addplot[color={rgb,405:red,209;green,216;blue,0}, error bars/.cd,y dir=both,y explicit]
            table[x=atom-loss-rate-improvement-factor, y=5 -successful-shots-before-reload 1, y error=5 -successful-shots-before-reload 1-err, col sep=comma]
            {data/hole-sensitivity-cnu-successful-shot-rate-sensitivity-cnu.csv}
        ;
        \addlegendentry{5};

        \addplot[color={rgb,405:red,0;green,205;blue,110}, error bars/.cd,y dir=both,y explicit]
            table[x=atom-loss-rate-improvement-factor, y=4 -successful-shots-before-reload 2, y error=4 -successful-shots-before-reload 2-err, col sep=comma]
            {data/hole-sensitivity-cnu-successful-shot-rate-sensitivity-cnu.csv}
        ;
        \addlegendentry{4};

        \addplot[color={rgb,405:red,125;green,0;blue,255}, error bars/.cd,y dir=both,y explicit]
            table[x=atom-loss-rate-improvement-factor, y=3 -successful-shots-before-reload 3, y error=3 -successful-shots-before-reload 3-err, col sep=comma]
            {data/hole-sensitivity-cnu-successful-shot-rate-sensitivity-cnu.csv}
        ;
        \addlegendentry{3};

    \end{loglogaxis}

\end{tikzpicture}%
        \hfill%
    }}
    \caption{Sensitivity to the rate of atom loss for the balanced \textit{Compile Small and Reroute} strategy. In prior experiments we used a fixed rate of 2\% atom loss. For larger systems this rate could be worse and in the future we might expect this rate to be much better. For each interaction distance we see as the rate of atom loss gets better we can run many more trials before we must perform a reload and reset. Some error bars don't show on the log axis.}
    \label{fig:hole-sensitivity}
\end{figure}
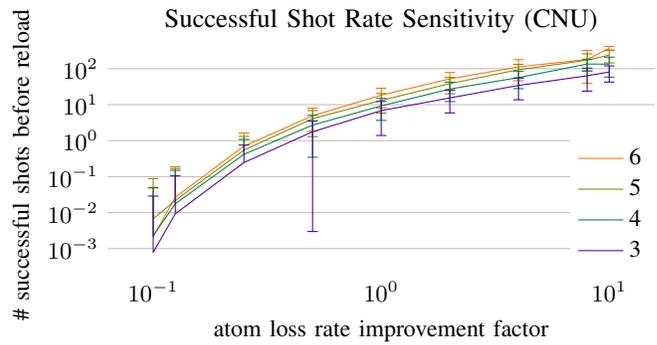
    
\begin{figure}
    \centering
    \scalebox{0.769}{
        \def\svgwidth{1.3\columnwidth}
        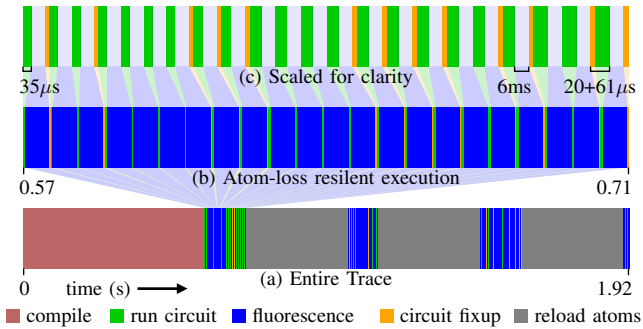%
    }
    \caption{A timeline of 20 successful shots for \textit{Compile Small and Reroute} with reload time of 0.3 s and fluorescing time of 6 ms. A majority of the overhead time is contributed by the reload time and fluorescence, indicating, that the duration and count of these actions is crucial to overall runtime.}
    \label{fig:run-trace}
\end{figure}

Recompilation is not shown in Figure \ref{fig:hole-timing} as software compilation exceeds the array reload time, and the overhead time is larger than simply reloading. Other strategies are always more time efficient than reloading all of the atoms.  Additionally, since compiling to a small size requires less fixes with swaps, the overhead time tends to be smaller at lower interaction distances.  As interaction distances increase, the overhead time of each strategy converges.  A sample timeline of 20 successful shots using compile small and reroute can be seen in Figure \ref{fig:run-trace}.  After initial compilation, reloading takes a majority of the time, so any ability to reduce the number of reloads vastly reduces the overall run time.

\textit{Compile small + reroute} is an efficient way to improve loss resilience, we next examine the sensitivity of the successful shot count before a reload to the rate of atom loss for this strategy.  Figure \ref{fig:hole-sensitivity} shows how the number of successful shots changes as the rate of atom loss changes.  A 10x improvement offers an expected 10x improvement in the number of successful shots before a reload must occur.  This is because the rate of atom loss decreases as technology improves, reducing the number of reloads, improving overhead time.

\section{Discussion and Conclusion}
Reducing the overhead of running compiled quantum programs is critical to successfully executing useful near- and intermediate term quantum algorithms.  Neutral atoms have many attractive properties: long-range interactions, native implementation of multiqubit gates, and ease of scalability.  These advantages reduce gate counts and depths dramatically in compiled circuits by increasing the relative connectivity of the underlying hardware. While long range interactions induce larger restriction zones which inhibit some parallelism, the amount of gate and depth savings far outweighs this cost. 

The dominant cost in NA systems is atom loss. Weaker trapping potential and destructive measurement leads to the loss of atoms as computation is performed.  We explore various strategies to adapt to this loss including the extremes of full recompilation and always reloading. Full recompilation is able to sustain high atom loss but is slow when thousands of trials are needed. But reloading is also slow and is the dominant hardware cost. Our reroute and compile small strategies balance atom loss resilience and shot success rate to save computation time. 

Popular competitor qubits have a head start on neutral atoms in terms of error rate and device sizes.  In simulation, we have demonstrated our large gate count and depth savings give advantage over superconducting systems. SC systems are often easy to increase in size, but fabrication variability and limited connectivity limit their effectiveness. Trapped-ion systems offer many of the same advantages as neutral atoms such as global interactions and multiqubit gates but at the cost of parallelism.  Ions also have stronger trapping potential, mitigating loss. Unfortunately, trapped ion systems will struggle to scale and maintain these properties. These systems have limited parallelism and are held in one dimensional traps limited to around 100 qubits. To scale, systems connect multiple traps with higher inter-trap communication cost \cite{Murali-ISCA20}. Neutral atom systems are theoretically capable of maintaining their advantages as they scale.

Even at a small scale, the unique properties of the NA systems result in compiled circuits which are lower depth and use fewer communication operations translating to an expected higher probability of success. These advantages will become even clearer as devices scale since the device connectivity per size grows much more favorably than other architectures. Our algorithms for compilation are scalable heuristics and will be able to keep up with increasing hardware size well. For atom loss, some techniques will not be favorable for larger device and program sizes, such as full recompilation, however we've shown other more clever and faster techniques are better suited for the problem and will be able to scale. For example, since the speed to adjusting a hardware mapping is on the order of nanoseconds, rather than the microseconds required to perform reloading and fluorescence, we can expect these techniques to remain viable.

In this work we have focused on software and hardware techniques to demonstrate neutral atoms, with our methods, are a viable and scalable alternative to more established technologies. Long-distance interactions and multiqubit operations dramatically reduce communication and depth overheads which translates into lower error rate requirements to obtain successful programs.  Like their competitors, there are fundamental drawbacks of a NA system; here we've highlighted the problem of atom loss. This probabilistic loss is inherent in the trapping process itself and prior hardware studies have focused hardware solutions to reduce this probability of loss. We demonstrate that software solutions can effectively mitigate the problems due to atom loss. This is critical for the overall development of the platform: by solving fundamental problems at the systems level, hardware developers can focus on solving and optimizing other problems and process of co-design which can accelerate the advancement of the hardware tremendously.

\bibliographystyle{IEEEtranS}
\bibliography{refs}

\begin{thebibliography}{10}
\providecommand{\url}[1]{#1}
\csname url@samestyle\endcsname
\providecommand{\newblock}{\relax}
\providecommand{\bibinfo}[2]{#2}
\providecommand{\BIBentrySTDinterwordspacing}{\spaceskip=0pt\relax}
\providecommand{\BIBentryALTinterwordstretchfactor}{4}
\providecommand{\BIBentryALTinterwordspacing}{\spaceskip=\fontdimen2\font plus
\BIBentryALTinterwordstretchfactor\fontdimen3\font minus
  \fontdimen4\font\relax}
\providecommand{\BIBforeignlanguage}[2]{{%
\expandafter\ifx\csname l@#1\endcsname\relax
\typeout{** WARNING: IEEEtranS.bst: No hyphenation pattern has been}%
\typeout{** loaded for the language `#1'. Using the pattern for}%
\typeout{** the default language instead.}%
\else
\language=\csname l@#1\endcsname
\fi
#2}}
\providecommand{\BIBdecl}{\relax}
\BIBdecl

\bibitem{github-link}
\BIBentryALTinterwordspacing
``Neutral atom compilation.'' [Online]. Available:
  \url{https://github.com/AndrewLitteken/neutral-atom-compilation}
\BIBentrySTDinterwordspacing

\bibitem{bristlecone}
\BIBentryALTinterwordspacing
``A preview of bristlecone, google's new quantum processor,'' Mar 2018.
  [Online]. Available:
  \url{https://ai.googleblog.com/2018/03/a-preview-of-bristlecone-googles-new.html}
\BIBentrySTDinterwordspacing

\bibitem{qiskit}
H.~Abraham, AduOffei, R.~Agarwal, I.~Y. Akhalwaya, G.~Aleksandrowicz,
  T.~Alexander, M.~Amy, E.~Arbel, Arijit02, A.~Asfaw, A.~Avkhadiev,
  C.~Azaustre, AzizNgoueya, A.~Banerjee, A.~Bansal, P.~Barkoutsos, A.~Barnawal,
  G.~Barron, G.~S. Barron, L.~Bello, Y.~Ben-Haim, D.~Bevenius, A.~Bhobe, L.~S.
  Bishop, C.~Blank, S.~Bolos, S.~Bosch, Brandon, S.~Bravyi, Bryce-Fuller,
  D.~Bucher, A.~Burov, F.~Cabrera, P.~Calpin, L.~Capelluto, J.~Carballo,
  G.~Carrascal, A.~Chen, C.-F. Chen, E.~Chen, J.~C. Chen, R.~Chen, J.~M. Chow,
  S.~Churchill, C.~Claus, C.~Clauss, R.~Cocking, F.~Correa, A.~J. Cross, A.~W.
  Cross, S.~Cross, J.~Cruz-Benito, C.~Culver, A.~D. C{\'o}rcoles-Gonzales,
  S.~Dague, T.~E. Dandachi, M.~Daniels, M.~Dartiailh, DavideFrr, A.~R. Davila,
  A.~Dekusar, D.~Ding, J.~Doi, E.~Drechsler, Drew, E.~Dumitrescu, K.~Dumon,
  I.~Duran, K.~EL-Safty, E.~Eastman, G.~Eberle, P.~Eendebak, D.~Egger,
  M.~Everitt, P.~M. Fern{\'a}ndez, A.~H. Ferrera, R.~Fouilland,
  FranckChevallier, A.~Frisch, A.~Fuhrer, B.~Fuller, M.~GEORGE, J.~Gacon, B.~G.
  Gago, C.~Gambella, J.~M. Gambetta, A.~Gammanpila, L.~Garcia, T.~Garg,
  S.~Garion, A.~Gilliam, A.~Giridharan, J.~Gomez-Mosquera, Gonzalo, S.~de~la
  Puente~Gonz{\'a}lez, J.~Gorzinski, I.~Gould, D.~Greenberg, D.~Grinko,
  W.~Guan, J.~A. Gunnels, M.~Haglund, I.~Haide, I.~Hamamura, O.~C. Hamido,
  F.~Harkins, V.~Havlicek, J.~Hellmers, {\L}.~Herok, S.~Hillmich, H.~Horii,
  C.~Howington, S.~Hu, W.~Hu, J.~Huang, R.~Huisman, H.~Imai, T.~Imamichi,
  K.~Ishizaki, R.~Iten, T.~Itoko, JamesSeaward, A.~Javadi, A.~Javadi-Abhari,
  W.~Javed, Jessica, M.~Jivrajani, K.~Johns, S.~Johnstun, Jonathan-Shoemaker,
  V.~K, T.~Kachmann, A.~Kale, N.~Kanazawa, Kang-Bae, A.~Karazeev, P.~Kassebaum,
  J.~Kelso, S.~King, Knabberjoe, Y.~Kobayashi, A.~Kovyrshin, R.~Krishnakumar,
  V.~Krishnan, K.~Krsulich, P.~Kumkar, G.~Kus, R.~LaRose, E.~Lacal, R.~Lambert,
  J.~Lapeyre, J.~Latone, S.~Lawrence, C.~Lee, G.~Li, D.~Liu, P.~Liu, Y.~Maeng,
  K.~Majmudar, A.~Malyshev, J.~Manela, J.~Marecek, M.~Marques, D.~Maslov,
  D.~Mathews, A.~Matsuo, D.~T. McClure, C.~McGarry, D.~McKay, D.~McPherson,
  S.~Meesala, T.~Metcalfe, M.~Mevissen, A.~Meyer, A.~Mezzacapo, R.~Midha,
  Z.~Minev, A.~Mitchell, N.~Moll, J.~Montanez, G.~Monteiro, M.~D. Mooring,
  R.~Morales, N.~Moran, M.~Motta, MrF, P.~Murali, J.~M{\"u}ggenburg,
  D.~Nadlinger, K.~Nakanishi, G.~Nannicini, P.~Nation, E.~Navarro, Y.~Naveh,
  S.~W. Neagle, P.~Neuweiler, J.~Nicander, P.~Niroula, H.~Norlen, NuoWenLei,
  L.~J. O'Riordan, O.~Ogunbayo, P.~Ollitrault, R.~Otaolea, S.~Oud, D.~Padilha,
  H.~Paik, S.~Pal, Y.~Pang, V.~R. Pascuzzi, S.~Perriello, A.~Phan, F.~Piro,
  M.~Pistoia, C.~Piveteau, P.~Pocreau, A.~Pozas-iKerstjens, M.~Prokop,
  V.~Prutyanov, D.~Puzzuoli, J.~P{\'e}rez, Quintiii, R.~I. Rahman, A.~Raja,
  N.~Ramagiri, A.~Rao, R.~Raymond, R.~M.-C. Redondo, M.~Reuter, J.~Rice,
  M.~Riedemann, M.~L. Rocca, D.~M. Rodr{\'\i}guez, RohithKarur, M.~Rossmannek,
  M.~Ryu, T.~SAPV, SamFerracin, M.~Sandberg, H.~Sandesara, R.~Sapra,
  H.~Sargsyan, A.~Sarkar, N.~Sathaye, B.~Schmitt, C.~Schnabel, Z.~Schoenfeld,
  T.~L. Scholten, E.~Schoute, J.~Schwarm, I.~F. Sertage, K.~Setia, N.~Shammah,
  Y.~Shi, A.~Silva, A.~Simonetto, N.~Singstock, Y.~Siraichi, I.~Sitdikov,
  S.~Sivarajah, M.~B. Sletfjerding, J.~A. Smolin, M.~Soeken, I.~O. Sokolov,
  I.~Sokolov, SooluThomas, Starfish, D.~Steenken, M.~Stypulkoski, S.~Sun, K.~J.
  Sung, H.~Takahashi, T.~Takawale, I.~Tavernelli, C.~Taylor, P.~Taylour,
  S.~Thomas, M.~Tillet, M.~Tod, M.~Tomasik, E.~de~la Torre, K.~Trabing,
  M.~Treinish, TrishaPe, D.~Tulsi, W.~Turner, Y.~Vaknin, C.~R. Valcarce,
  F.~Varchon, A.~C. Vazquez, V.~Villar, D.~Vogt-Lee, C.~Vuillot, J.~Weaver,
  J.~Weidenfeller, R.~Wieczorek, J.~A. Wildstrom, E.~Winston, J.~J. Woehr,
  S.~Woerner, R.~Woo, C.~J. Wood, R.~Wood, S.~Wood, S.~Wood, J.~Wootton,
  D.~Yeralin, D.~Yonge-Mallo, R.~Young, J.~Yu, C.~Zachow, L.~Zdanski, H.~Zhang,
  C.~Zoufal, Zoufalc, a~kapila, a~matsuo, bcamorrison, brandhsn, nick bronn,
  brosand, chlorophyll zz, csseifms, dekel.meirom, dekelmeirom, dekool, dime10,
  drholmie, dtrenev, ehchen, elfrocampeador, faisaldebouni, fanizzamarco,
  gabrieleagl, gadial, galeinston, georgios ts, gruu, hhorii, hykavitha,
  jagunther, jliu45, jscott2, kanejess, klinvill, krutik2966, kurarrr,
  lerongil, ma5x, merav aharoni, michelle4654, ordmoj, sagar pahwa, rmoyard,
  saswati qiskit, scottkelso, sethmerkel, shaashwat, sternparky, strickroman,
  sumitpuri, tigerjack, toural, tsura crisaldo, vvilpas, welien, willhbang,
  yang.luh, yotamvakninibm, and M.~{\v{C}}epulkovskis, ``Qiskit: An open-source
  framework for quantum computing,'' 2019.

\bibitem{look2}
J.~M. Baker, C.~Duckering, A.~Hoover, and F.~T. Chong, ``Time-sliced quantum
  circuit partitioning for modular architectures,'' in \emph{Proceedings of the
  17th ACM International Conference on Computing Frontiers}, 2020, pp. 98--107.

\bibitem{barenco}
A.~Barenco, C.~H. Bennett, R.~Cleve, D.~P. DiVincenzo, N.~Margolus, P.~Shor,
  T.~Sleator, J.~A. Smolin, and H.~Weinfurter, ``Elementary gates for quantum
  computation,'' \emph{Physical review A}, vol.~52, no.~5, p. 3457, 1995.

\bibitem{na3}
D.~Barredo, S.~De~L{\'e}s{\'e}leuc, V.~Lienhard, T.~Lahaye, and A.~Browaeys,
  ``An atom-by-atom assembler of defect-free arbitrary two-dimensional atomic
  arrays,'' \emph{Science}, vol. 354, no. 6315, pp. 1021--1023, 2016.

\bibitem{na5}
D.~Barredo, V.~Lienhard, S.~De~Leseleuc, T.~Lahaye, and A.~Browaeys,
  ``Synthetic three-dimensional atomic structures assembled atom by atom,''
  \emph{Nature}, vol. 561, no. 7721, pp. 79--82, 2018.

\bibitem{bv}
E.~Bernstein and U.~Vazirani, ``Quantum complexity theory,'' \emph{SIAM Journal
  on computing}, vol.~26, no.~5, pp. 1411--1473, 1997.

\bibitem{ion-scale}
C.~D. Bruzewicz, J.~Chiaverini, R.~McConnell, and J.~M. Sage, ``Trapped-ion
  quantum computing: Progress and challenges,'' \emph{Applied Physics Reviews},
  vol.~6, no.~2, p. 021314, 2019.

\bibitem{vacuum-atom-loss}
\BIBentryALTinterwordspacing
J.~P. Covey, I.~S. Madjarov, A.~Cooper, and M.~Endres, ``2000-times repeated
  imaging of strontium atoms in clock-magic tweezer arrays,'' \emph{Phys. Rev.
  Lett.}, vol. 122, p. 173201, May 2019. [Online]. Available:
  \url{https://link.aps.org/doi/10.1103/PhysRevLett.122.173201}
\BIBentrySTDinterwordspacing

\bibitem{routing1}
A.~Cowtan, S.~Dilkes, R.~Duncan, A.~Krajenbrink, W.~Simmons, and S.~Sivarajah,
  ``On the qubit routing problem,'' \emph{arXiv preprint arXiv:1902.08091},
  2019.

\bibitem{cuccaro}
S.~A. Cuccaro, T.~G. Draper, S.~A. Kutin, and D.~P. Moulton, ``A new quantum
  ripple-carry addition circuit,'' \emph{arXiv preprint quant-ph/0410184},
  2004.

\bibitem{dram-time}
V.~{Cuppu}, B.~{Jacob}, B.~{Davis}, and T.~{Mudge}, ``A performance comparison
  of contemporary {DRAM} architectures,'' in \emph{Proceedings of the 26th
  International Symposium on Computer Architecture (Cat. No.99CB36367)}, 1999,
  pp. 222--233.

\bibitem{na2}
M.~Endres, H.~Bernien, A.~Keesling, H.~Levine, E.~R. Anschuetz, A.~Krajenbrink,
  C.~Senko, V.~Vuletic, M.~Greiner, and M.~D. Lukin, ``Atom-by-atom assembly of
  defect-free one-dimensional cold atom arrays,'' \emph{Science}, vol. 354, no.
  6315, pp. 1024--1027, 2016.

\bibitem{qaoa}
E.~Farhi, J.~Goldstone, and S.~Gutmann, ``A quantum approximate optimization
  algorithm,'' \emph{arXiv preprint arXiv:1411.4028}, 2014.

\bibitem{meas-loss-1}
\BIBentryALTinterwordspacing
A.~Fuhrmanek, R.~Bourgain, Y.~R.~P. Sortais, and A.~Browaeys, ``Free-space
  lossless state detection of a single trapped atom,'' \emph{Phys. Rev. Lett.},
  vol. 106, p. 133003, Mar 2011. [Online]. Available:
  \url{https://link.aps.org/doi/10.1103/PhysRevLett.106.133003}
\BIBentrySTDinterwordspacing

\bibitem{gottesman2010introduction}
D.~Gottesman, ``An introduction to quantum error correction and fault-tolerant
  quantum computation,'' in \emph{Quantum information science and its
  contributions to mathematics, Proceedings of Symposia in Applied
  Mathematics}, vol.~68, 2010, pp. 13--58.

\bibitem{grover}
L.~K. Grover, ``A fast quantum mechanical algorithm for database search,'' in
  \emph{Proceedings of the Twenty-Eighth Annual ACM Symposium on Theory of
  Computing}, 1996, pp. 212--219.

\bibitem{scheduling1}
G.~G. Guerreschi and J.~Park, ``Two-step approach to scheduling quantum
  circuits,'' \emph{Quantum Science and Technology}, vol.~3, no.~4, p. 045003,
  2018.

\bibitem{Henriet_2020}
\BIBentryALTinterwordspacing
L.~Henriet, L.~Beguin, A.~Signoles, T.~Lahaye, A.~Browaeys, G.-O. Reymond, and
  C.~Jurczak, ``Quantum computing with neutral atoms,'' \emph{Quantum}, vol.~4,
  p. 327, Sep 2020. [Online]. Available:
  \url{http://dx.doi.org/10.22331/q-2020-09-21-327}
\BIBentrySTDinterwordspacing

\bibitem{routing2}
Y.~Hirata, M.~Nakanishi, S.~Yamashita, and Y.~Nakashima, ``An efficient
  conversion of quantum circuits to a linear nearest neighbor architecture,''
  \emph{Quantum Information and Computation}, vol.~11, no.~1, p. 142, 2011.

\bibitem{ibmq}
``{IBM Quantum Devices},''
  \url{https://quantumexperience.ng.bluemix.net/qx/devices}, accessed:
  2020-11-24.

\bibitem{na-gates}
D.~Jaksch, J.~Cirac, P.~Zoller, S.~Rolston, R.~C{\^o}t{\'e}, and M.~Lukin,
  ``Fast quantum gates for neutral atoms,'' \emph{Physical Review Letters},
  vol.~85, no.~10, p. 2208, 2000.

\bibitem{look4}
\BIBentryALTinterwordspacing
S.~Jandura, ``Improving a quantum compiler,'' Sep 2018. [Online]. Available:
  \url{https://medium.com/qiskit/improving-a-quantum-compiler-48410d7a7084}
\BIBentrySTDinterwordspacing

\bibitem{na4}
H.~Kim, W.~Lee, H.-g. Lee, H.~Jo, Y.~Song, and J.~Ahn, ``In situ single-atom
  array synthesis using dynamic holographic optical tweezers,'' \emph{Nature
  communications}, vol.~7, no.~1, pp. 1--8, 2016.

\bibitem{oliver1}
M.~Kjaergaard, M.~E. Schwartz, J.~Braum{\"u}ller, P.~Krantz, J.~I.-J. Wang,
  S.~Gustavsson, and W.~D. Oliver, ``Superconducting qubits: Current state of
  play,'' \emph{Annual Review of Condensed Matter Physics}, vol.~11, pp.
  369--395, 2020.

\bibitem{meas-loss-2}
\BIBentryALTinterwordspacing
M.~Kwon, M.~F. Ebert, T.~G. Walker, and M.~Saffman, ``Parallel low-loss
  measurement of multiple atomic qubits,'' \emph{Phys. Rev. Lett.}, vol. 119,
  p. 180504, Oct 2017. [Online]. Available:
  \url{https://link.aps.org/doi/10.1103/PhysRevLett.119.180504}
\BIBentrySTDinterwordspacing

\bibitem{multiqubitgates}
H.~Levine, A.~Keesling, G.~Semeghini, A.~Omran, T.~T. Wang, S.~Ebadi,
  H.~Bernien, M.~Greiner, V.~Vuleti{\'c}, H.~Pichler \emph{et~al.}, ``Parallel
  implementation of high-fidelity multiqubit gates with neutral atoms,''
  \emph{Physical review letters}, vol. 123, no.~17, p. 170503, 2019.

\bibitem{map1}
P.~Murali, J.~M. Baker, A.~Javadi-Abhari, F.~T. Chong, and M.~Martonosi,
  ``Noise-adaptive compiler mappings for noisy intermediate-scale quantum
  computers,'' in \emph{Proceedings of the Twenty-Fourth International
  Conference on Architectural Support for Programming Languages and Operating
  Systems}, 2019, pp. 1015--1029.

\bibitem{Murali-ISCA20}
\BIBentryALTinterwordspacing
P.~Murali, D.~M. Debroy, K.~R. Brown, and M.~Martonosi, \emph{Architecting
  Noisy Intermediate-Scale Trapped Ion Quantum Computers}.\hskip 1em plus 0.5em
  minus 0.4em\relax IEEE Press, 2020, p. 529–542. [Online]. Available:
  \url{https://doi.org/10.1109/ISCA45697.2020.00051}
\BIBentrySTDinterwordspacing

\bibitem{xtalk}
P.~Murali, D.~C. McKay, M.~Martonosi, and A.~Javadi-Abhari, ``Software
  mitigation of crosstalk on noisy intermediate-scale quantum computers,'' in
  \emph{Proceedings of the Twenty-Fifth International Conference on
  Architectural Support for Programming Languages and Operating Systems}, 2020,
  pp. 1001--1016.

\bibitem{dram-spare}
P.~J. Nair, ``Architectural techniques to enable reliable and scalable memory
  systems,'' \emph{arXiv preprint arXiv:1704.03991}, 2017.

\bibitem{cancel}
Y.~Nam, N.~J. Ross, Y.~Su, A.~M. Childs, and D.~Maslov, ``Automated
  optimization of large quantum circuits with continuous parameters,''
  \emph{npj Quantum Information}, vol.~4, no.~1, pp. 1--12, 2018.

\bibitem{mikeike}
M.~A. Nielsen and I.~L. Chuang, \emph{Quantum Computation and Quantum
  Information: 10th Anniversary Edition}, 10th~ed.\hskip 1em plus 0.5em minus
  0.4em\relax New York, NY, USA: Cambridge University Press, 2011.

\bibitem{Ohl_de_Mello_2019}
\BIBentryALTinterwordspacing
D.~Ohl~de Mello, D.~Schäffner, J.~Werkmann, T.~Preuschoff, L.~Kohfahl,
  M.~Schlosser, and G.~Birkl, ``Defect-free assembly of {2D} clusters of more
  than 100 single-atom quantum systems,'' \emph{Physical Review Letters}, vol.
  122, no.~20, May 2019. [Online]. Available:
  \url{http://dx.doi.org/10.1103/PhysRevLett.122.203601}
\BIBentrySTDinterwordspacing

\bibitem{qft-adder}
L.~Ruiz-Perez and J.~C. Garcia-Escartin, ``Quantum arithmetic with the quantum
  fourier transform,'' \emph{Quantum Information Processing}, vol.~16, no.~6,
  p. 152, 2017.

\bibitem{saffman}
M.~Saffman, ``Quantum computing with atomic qubits and rydberg interactions:
  progress and challenges,'' \emph{Journal of Physics B: Atomic, Molecular and
  Optical Physics}, vol.~49, no.~20, p. 202001, 2016.

\bibitem{shor}
P.~W. Shor, ``Polynomial-time algorithms for prime factorization and discrete
  logarithms on a quantum computer,'' \emph{SIAM Journal on Computing}, 1995.

\bibitem{rigetti}
R.~S. Smith, M.~J. Curtis, and W.~J. Zeng, ``A practical quantum instruction
  set architecture,'' \emph{arXiv preprint arXiv:1608.03355}, 2016.

\bibitem{map2}
S.~S. Tannu and M.~Qureshi, ``Ensemble of diverse mappings: Improving
  reliability of quantum computers by orchestrating dissimilar mistakes,'' in
  \emph{Proceedings of the 52nd Annual IEEE/ACM International Symposium on
  Microarchitecture}, 2019, pp. 253--265.

\bibitem{python}
G.~Van~Rossum and F.~L. Drake, \emph{Python 3 Reference Manual}.\hskip 1em plus
  0.5em minus 0.4em\relax Scotts Valley, CA: CreateSpace, 2009.

\bibitem{map3}
R.~Wille, L.~Burgholzer, and A.~Zulehner, ``Mapping quantum circuits to {IBM}
  {QX} architectures using the minimal number of {SWAP} and {H} operations,''
  in \emph{2019 56th ACM/IEEE Design Automation Conference (DAC)}.\hskip 1em
  plus 0.5em minus 0.4em\relax IEEE, 2019, pp. 1--6.

\bibitem{look1}
R.~Wille, O.~Keszocze, M.~Walter, P.~Rohrs, A.~Chattopadhyay, and R.~Drechsler,
  ``Look-ahead schemes for nearest neighbor optimization of {1D} and {2D}
  quantum circuits,'' in \emph{2016 21st Asia and South Pacific design
  automation conference (ASP-DAC)}.\hskip 1em plus 0.5em minus 0.4em\relax
  IEEE, 2016, pp. 292--297.

\bibitem{ionq}
K.~Wright, K.~Beck, S.~Debnath, J.~Amini, Y.~Nam, N.~Grzesiak, J.-S. Chen,
  N.~Pisenti, M.~Chmielewski, C.~Collins \emph{et~al.}, ``Benchmarking an
  11-qubit quantum computer,'' \emph{Nature communications}, vol.~10, no.~1,
  pp. 1--6, 2019.

\end{thebibliography}

\end{document}